\documentclass[aps,showpacs,preprint,epsf,epsfig, nofootinbib]{revtex4-2}
\usepackage{mathrsfs}
\usepackage[english]{babel}[2020/02/03]
\makeatletter\AtBeginDocument{\let\@elt\relax}\makeatother
\usepackage{dcolumn}% Align table columns on decimal point
\usepackage{bm}% bold math
\usepackage{hyperref}% add hypertext capabilities
\usepackage{epsfig,graphicx,color,appendix}
\usepackage{multirow}
\usepackage{tabularx}
\usepackage{capt-of}
\usepackage{caption}
\usepackage[countmax]{subfloat}

\usepackage{threeparttable}
\usepackage{amssymb}
\usepackage{array}
\usepackage{amsmath}
\usepackage{rotating,tabularx}
\usepackage{float}
\usepackage{xcolor}
\usepackage[english]{babel}

\begin{document}
	\def\ben{\begin{eqnarray}}
		\def\en{\end{eqnarray}}
	\def\t{\times}
	\def\pp{{\prime\prime}}
	\def\nc{N_c^{\rm eff}}
	\def\vp{\varepsilon}
	\def\hep{\hat{\varepsilon}}
	\def\e{{\cal E}}
	\def\up{\uparrow}
	\def\dw{\downarrow}
	\long\def\symbolfootnote[#1]#2{\begingroup%
		\def\thefootnote{\fnsymbol{footnote}}\footnote[#1]{#2}\endgroup}
	\def\lsim{ {\ \lower-1.2pt\vbox{\hbox{\rlap{$<$}\lower5pt\vbox{\hbox{$\sim$}
			}}}\ } }
	\def\gsim{ {\ \lower-1.2pt\vbox{\hbox{\rlap{$>$}\lower5pt\vbox{\hbox{$\sim$}
			}}}\ } }
	
	\font\el=cmbx10 scaled \magstep2{\obeylines\hfill \today}
	\vskip 1.5 cm
	\centerline{\large\bf Screening of quark charge and mixing effects on}
	\centerline{\large\bf  transition moments and M1 decay widths of baryons}
	\small
	\vskip 1.0 cm
	
	\centerline{\bf Binesh Mohan\footnote[1]{\href{mailto:bineshmohan96@gmail.com}{bineshmohan96@gmail.com}}, Thejus Mary S.\footnote[2]{\href{mailto:96theju13@gmail.com}{96theju13@gmail.com}}, Avijit Hazra\footnote[3]{\href{mailto:avijithr@srmist.edu.in}{avijithr@srmist.edu.in}}, and Rohit Dhir\footnote[4]{Corresponding author: \href{mailto:dhir.rohit@gmail.com}{dhir.rohit@gmail.com}}}
	
	\medskip
	\centerline{\it Department of Physics and Nanotechnology,}
	\centerline{\it SRM Institute of Science and Technology, Kattankulathur 603203, India.}
	\bigskip
	\bigskip
	\begin{center}
		{\large \bf Abstract}
	\end{center}
	Motivated by the precision measurements of heavy flavor baryon masses, we analyze the modification of quark charge by employing the screening effect inside the baryon. In addition, we calculate the isospin mass splitting up to charmed baryons employing isospin symmetry breaking. Consequently, we obtain the masses, magnetic moments, and transition moments of $J^P=\frac{1}{2}^+$ and $\frac{3}{2}^+$ baryons to predict radiative decay widths for $\frac{1}{2}^{\prime +} \to \frac{1}{2}^+$ and $\frac{3}{2}^+\to \frac{1}{2}^{(\prime)+}$ transitions. Finally, we include the effects of state mixing in flavor degenerate baryon magnetic and transition moments, as well as M1 transition decay widths. 
	
	\medskip
	\medskip
	
	\vfill
	
	\section {Introduction}
	\label{introduction}
	Most recently, the LHCb and CMS collaboration have reported the observation of excited bottom-strange states $\Xi_b(6327)^0$, $\Xi_b(6333)^0$, $\Xi_b(6100)^-$, and the isospin partner of $\Xi_b^{-}$, \textit{i.e.}, $\Xi_b(6227)^0$ \cite{LHCb:2021ssn, LHCb:2020xpu, CMS:2021rvl}. The two new charm states $\Xi_c(2923)^0$ and $\Xi_c(2939)^0$, an excited state $\Lambda_b(6070)^0$, and four narrow peaks for $\Omega_b^-$ excited states have also been reported \cite{LHCb:2020iby, Aaij:2020rkw, Aaij:2020cex}. In addition, the observations and precision measurements of all the $J^P=\frac{1}{2}^+ ~\text{and}~\frac{3}{2}^+$ singly heavy charmed baryons present a complete picture of the low-lying charm baryon spectroscopy \cite{Meng:2022ozq, Workman:2022ynf, LHCb:2014chk, LHCb:2019sxa, LHCb:2017zzt, LHCb:2016mrc, LHCb:2016coe, CDF:2014mon, LHCb:2014nae, LHCb:2013jih, Solovieva:2008fw, Chen:2016spr}. However, the masses of multiple bottom states, \textit{like}, $\Xi_b^{\prime0}$, $\Sigma_b^{0}$, $\Sigma_b^{*0}$, and $\Omega_b^{*-}$ are not confirmed as yet. The existence of $\Xi_{cc}^{++}$ state was confirmed by the LHCb collaboration in 2017, and the mass of its isoplet partner $\Xi_{cc}^{+}$ has been updated based on recent searches \cite{Aaij:2017ueg, LHCb:2019epo, LHCb:2019gqy,LHCb:2021eaf}. The LHCb is conducting new investigations for the doubly heavy baryons, $\Xi_{cb}^{0}$ and $\Omega_{cb}^{0}$; however, no significant signal has been found \cite{LHCb:2021xba, Aaij:2020vid}. Further measurements will be possible with larger data samples, and additional decay modes are expected at the upgraded LHCb experiments \cite{LHCb:2018roe}. On the other hand, the measurements of the heavy flavor magnetic moments need more experimental effort. Fomin \textit{et al.} \cite{Fomin:2019wuw} have discussed the prospects of magnetic moment measurements for charmed baryons by analyzing the radiative charmonium decay at BES III \cite{BESIII:2017tsq}, although the experimental growth in the assessment of electromagnetic properties of charmed baryons is still moderate. The radiative decay processes, namely, $\Omega_{c}^{*0}\rightarrow\Omega_{c}^{0}\gamma$, $\Xi_{c}^{\prime+}\rightarrow\Xi_{c}^{+}\gamma$, and $\Xi_c^{\prime0}\rightarrow\Xi_{c}^{0}\gamma$, were observed but not measured experimentally by the BaBar and Belle collaborations \cite{Cheng:2021qpd, Aubert:2006je, Solovieva:2008fw, Jessop:1998wt, BaBar:2006bjt, Yelton:2016fqw}. We expect more experimental results on the radiative decay widths of charm and bottom baryons in the near future from BES III and LHCb \cite{Fomin:2019wuw, Ablikim:2019hff, Yuan:2019zfo, Aiola:2020yam, Audurier:2021wqk}. 
	
	The enormous experimental activities have motivated a number of theorists and phenomenologists to explore the heavy flavor physics with improved interest. A number of theoretical studies on masses and electromagnetic properties of heavy flavor baryons have been conducted. These properties and decays are the keys to understanding heavy flavor structure and dynamics. In this context, heavy flavor spectroscopy, magnetic (transition) moments, and radiative decays provide the testing hypotheses for distinct models of hadronic structure. The effective mass scheme (EMS) and the screened quark charge effects have been successfully used for predicting the magnetic properties of charmed as well as bottom baryons \cite{Kumar:2005ei, Dhir:2009ax, Dhir:2013nka, Hazra:2021lpa}. The current work focuses on the predictions of magnetic (transition) moments and radiative M1 decay widths of $J^P=\frac{1}{2}^+ ~\text{and}~\frac{3}{2}^+$ charmed baryons in EMS with screened quark charge. However, we also present results in strange sector to refine our model against existing experimental results. In addition, we include isospin symmetry breaking through constituent quark masses and strong hyperfine interaction terms. We estimate the isospin symmetry breaking in different flavor sectors up to charm, which in turn determines the isospin mass splitting in baryons. Further, we integrate the screened quark charge effects on magnetic (transition) moments and M1 decay widths. Consequently, we predict the transition moments and M1 radiative decay widths of light and charm baryons corresponding to $B^{\prime} (\frac{1}{2}^{+})\to B(\frac{1}{2}^+)$ and $B^{*} (\frac{3}{2}^+)\to B^{(\prime)} (\frac{1}{2}^{+})$ transitions. Finally, we concentrate on the state mixing effects in our calculations. We believe that, though isospin symmetry breaking is minimal in magnetic moments, the screened charge and state mixing effects are of significant magnitude, which further improves our comparison with the available experimental results.
	
	The paper is organized as follows: Sections \ref{EQS}, \ref{isospin_splitting}, \ref{SQC} and \ref{properties} explain the necessary methodology. In Section \ref{results}, numerical results and their comparison with other theoretical predictions are provided, followed by a detailed analysis. In Section \ref{state_mixing}, we present the state mixing effects and analyze their impact on our results. We list our conclusion in the last section. Furthermore, we give various sum rules among the masses and magnetic moments in Appendices \ref{appendixA} and \ref{appendixB}, respectively.

	\section {Effective Quark Mass}
	\label{EQS}
	The mass of the quarks inside a baryon can be modified due to the one-gluon exchange interaction with the spectator quarks \cite{DeRujula:1975qlm} and is referred to as the effective quark mass in the concept of EMS. According to EMS, the baryon mass, $M_B$, is calculated as the sum of the effective masses of all the three quarks inside a baryon, and it can also be written as the sum of the constituent quark masses and the spin-dependent strong hyperfine interaction terms \cite{Verma:1987, Kumar:2005ei, Dhir:2009ax, Dhir:2013nka, Hazra:2021lpa}. Thus, 
	\begin{equation}
		\label{e1} 
		{M}_B = \sum_{i}m_{i} ^{\mathscr{E}} = \sum_{i}{m_i} + \sum_{i<j}b_{ij}\bf{s_i.s_j},
	\end{equation}
	where, $m_{i}^{\mathscr{E}}$ and $m_{i}$ represent the effective and constituent masses of the quark, $i$, within the baryon, respectively; and the spin operators of the $i^{th}$ and $j^{th}$ quarks are denoted by $\bf{s_i}$ and $\bf{s_j}$, respectively. The strong hyperfine interaction term\footnote{Note that, $b_{ij}=b_{ji}$.}, $b_{ij}$, for baryon $B(ijk)$ is given by
	\begin{equation}
		\label{e2}
		{b}_{ij} =\frac{16\pi\alpha_s}{9m_im_j}<\psi|\delta^3(\vec{r})|\psi>,
	\end{equation}
	where, $\psi$ is the baryon wave function at the origin and $\alpha_s$ is the strong coupling constant. The mass of a quark inside a baryon, $B(ijk)$, changes due to interactions with other quarks. Therefore, general expressions for baryon masses can be written in terms of effective quark masses as \footnote{Follow Appendix \ref{bar_mass_exp} for more details on effective mass relations.}:
	\begin{enumerate}	
		\item[1.] For octet baryons,
		\begin{enumerate}
			\item[(a)] Anti-symmetric $\Lambda_{[i j]k}-$type ($J^P=\frac{1}{2}^+$) baryons have,
			\begin{equation}
				\label{b1}
				{M}_B  = m_{i} + m_{j} + m_{k} - \frac{3b_{i j}}{4}. 
			\end{equation}
			\item[(b)] Symmetric $\Sigma_{\{i j\}k}-$type ($J^P=\frac{1}{2}^{\prime+}$) baryons takes the form,
			\begin{equation}
				\label{b2}
				{M}_{B^{\prime}}  = m_{i} + m_{j} + m_{k} + \frac{b_{i j}}{4} - \frac{b_{ jk}}{2} - \frac{b_{i k}}{2}.
			\end{equation}
		\end{enumerate}
		\item[2.] For decuplet ($J^P=\frac{3}{2}^+$) baryons,
		\begin{equation}
			\label{b3}
			{M}_{B^{*}}  = m_{i} + m_{j} + m_{k} + \frac{b_{i j}}{4} + \frac{b_{ jk}}{4} + \frac{b_{i k}}{4},
		\end{equation}
	\end{enumerate}
	where, $i$, $j$, and $k$ corresponds to the first, second, and third quarks, respectively. The effective quark mass, $m_i^{\mathscr{E}}$, defined in EMS (see Appendix \ref{bar_mass_exp}) is equivalent to the leading order parametrization of the baryon mass in chiral perturbation theory ($\chi$PT) \cite{Morpurgo:1989my, Dillon:1995qw}. The effective mass parametrization goes beyond the leading order in quark mass splitting because of the $1/m_i m_j$ term appearing via the hyperfine interaction. Furthermore, it can be argued that more complex quantum chromodynamics (QCD) parametrization of baryon mass can be simplified for the reason that flavor breaking in Lagrangian is caused by quark mass difference, and electromagnetic charge is only carried by quarks \cite{Dillon:2002ks}. Moreover, the higher order nonlinear terms have decreasing coefficients, which is evident from the fact that the Gell-Mann–Okubo mass relation is fulfilled \cite{Durand:2001zz, Durand:2001sz,Dillon:2002ks}. Therefore, the nonrelativistic quark model (NRQM) calculations for masses and magnetic moments are completely equivalent to leading order parametrization of relativistic field theory. Thus, the general expressions for baryon masses given by (\ref{b1})-(\ref{b3}) include flavor breaking for all terms (in $m_i$ and $b_{ij}$) to first order and are sufficient to reproduce baryon masses. Thus, we use effective quark mass originating from the one-gluon exchange interaction to calculate the magnetic (transition) moments and, consequently, M1 decay widths.
	
	We wish to emphasize that EMS utilizes experimental information to calculate constituent quark masses and hyperfine interaction terms, $b_{ij}$, in a model-independent manner. Following the equations mentioned in Appendix \ref{bar_mass_exp}, we have calculated the constituent quark masses and hyperfine interaction terms individually for each flavor sector from known experimental masses as given in Table \ref{Quark mass isospin}. In addition, we have included isospin symmetry breaking to refine our calculations which is discussed in next section. Furthermore, to include flavor-dependent effects, the $b_{ij}$ terms are obtained from corresponding flavor sectors to produce reliable results. 
	%%%%%%%%%%%%%%%%%%%%%%%%%%%%%%%%%%%%%%%%%%%%%%%%%%%%%%%%%%%%%%%
	\begin{table}[ht]
		\centering
		\captionof{table}{Constituent quark masses and hyperfine interaction terms with isospin splitting \symbolfootnote[1]{In the current work, we have neglected the uncertainties unlike our previous work \cite{Hazra:2021lpa}, for being very small.} (in MeV).}
		\label{Quark mass isospin}
		\setlength{\tabcolsep}{2pt}
		\begin{tabular}{|c|c||c|c|} 	\hline 
			\textbf{Experimental}			&\textbf{Constituent Quark} & \textbf{Experimental} &\textbf{Hyperfine Interaction} \\
			\textbf{Inputs \cite{Workman:2022ynf}} &\textbf{Masses ($m_{i}$)} & \textbf{Inputs \cite{Workman:2022ynf} } &\textbf{Terms ($b_{ij}$)}\\	\hline \hline
			$N, N^{*}$\symbolfootnote[2]{The values are fixed through minimization using the experimental masses of all the baryons in the $SU(2)$ sector.} &$m_{u} = 360.534$ & $N$,$ N^{*}$   &  $b_{uu} = 200.536$\\
			&$m_{d} =  363.491$  &  $N$, $\Delta^{+} $                  &   $b_{ud} = 197.752$ \\
			&                    &  $N$, $ N^{*}$                        &   $b_{dd} = 193.884$ \\	 \hline
			$\Lambda^{0}$                       &$m_{s} =  539.972$  &  $\Xi^{0}$                         &   $b_{us} = 143.129$ \\
			&                    &  $\Xi^{-}$                         &   $b_{ds} = 139.236$ \\
			&                    &  $\Omega^{-}$                      &   $b_{ss} = 70.045$  \\	\hline	
			$\Omega_{c}^{0}$, $\Omega_{c}^{*0}$ &$m_{c} = 1644.878 $ &  $\Xi_{c}^{\prime+}$, $\Xi_{c}^{*+}$ &   $b_{uc} = 42.067$  \\
			&                    &  $\Xi_{c}^{\prime0}$, $\Xi_{c}^{*0}$ &   $b_{dc} = 42.814$  \\
			&                    &  $\Omega_{c}^{0}$                  &   $b_{sc} = 47.133 $ \\
			&                    &  $\Xi_{cc}^{++}$                   &   $b_{cc} = 53.508 $ \\	\hline 
		\end{tabular} 
	\end{table}   
	%%%%%%%%%%%%%%%%%%%%%%%%%%%%%%%%%%%%%%%%%%%%%%%%%%%%%%%%%%%%%%%%%% 
	
	\section{Isospin Splitting}
	\label{isospin_splitting}
	Isospin splitting distinguishes between the charged states of an isospin multiplet. The isospin mass splitting in baryons may arise from a combination of effects, including the intrinsic mass difference between $u$ and $d$ quarks $(m_{d} - m_{u})$, the electromagnetic and color hyperfine interactions between the neighboring quarks, and the pairwise Coulomb interactions \cite{DeRujula:1975qlm, Rosner:1998zc, Karliner:2019lau, Karliner:2008sv}. The intrinsic mass difference between the quarks suffer from contributions arising from the interactions with neighboring quarks. These contributions are not only difficult to calculate, but are also model-dependent. It has been seen in the past that isospin splittings in different models suffer from strong parameter dependence, which cannot be avoided \cite{Cutkosky:1993cc}. For accurate assessment of several small contributions, more precise experimental data is required. Furthermore, for baryons containing one heavy quark, the electromagnetic contributions are expected to be small, allowing $m_ d-m_ u$ and strong hyperfine interaction terms to dominate \cite{Isgur:1979ed}. Also, it has been pointed out that the electromagnetic hyperfine contributions to the isospin mass splitting do not exceed the systematic uncertainty in experimental results and can be neglected for baryons with one or more heavy quarks \cite{Rosner:2006yk, Hwang:2008dj}. Color (electromagnetic) hyperfine interaction contributions, on the other hand, are determined by quark masses (charges) and wave function overlap at the origin, $|\psi(0)|^2$. However, $\alpha_{s}$ and $|\psi(0)|^2$ (that appear in both electromagnetic and strong hyperfine interactions) are scale dependent and, in most models, are assumed to be same for all the baryons. In short, the involvement of multiple parameters makes evaluation of the isospin symmetry breaking contributions from various individual sources nontrivial. Moreover, isospin symmetry breaking, though important, is expected to be small in hadrons. Thus, involvement of too many parameters corresponding to each contribution does not present the clear picture of isospin symmetry breaking.
	
	It is well known that all the aforementioned contributions to the observed baryon masses, originating from different interactions, can be approximated by the renormalization of quark masses. The key feature of EMS is that the constituent quark masses include all the other contributions except strong hyperfine interactions. Moreover, EMS uniquely defines the underlying quark structure of the baryon independent of parameters, where all the hyperfine interaction ($b_{ij}$) terms are precisely determined from the measured experimental baryon masses \cite{Workman:2022ynf}. As compared to our previous work \cite{Hazra:2021lpa}, we focus on the description of isospin splitting in EMS by estimating constituent quark masses and hyperfine interaction terms from respective flavor sectors to establish a more realistic picture. Thus, in our results, the isospin splitting is incorporated through quark masses, $m_i$, and strong hyperfine interaction terms, $b_{ij}$, as shown in Table \ref{Quark mass isospin}. We have used $N~\text{and}~N^*$ as inputs to minimize the mass relations (given in Appendix \ref{bar_mass_exp}) of the baryons corresponding to available experimental values to obtain $m_u$, $m_d$, $b_{uu}$, and $b_{dd}$ in the $SU(2)$ symmetry using the package MINUIT \cite{James:1975dr}. It is worth noting that we obtained a $\chi^{2}$ value as low as 0.0000169 for the fit, as defined by, 
	
	\begin{equation} 
		\label{e4}
		\chi^{2} = \sum_{i} (\frac{X^{Th.}_{i} - X^{Expt.}_{i}}{X^{Expt.}_{i}})^{2},
	\end{equation}
	where, $X^{Th.}_{i}$ represent the theoretical masses given by (\ref{e1}), and $X^{Expt.}_{i}$ are experimental masses of the baryons. The hyperfine interaction term, $b_{ud}$, on the other hand, has been determined from experimental masses of $N$ and $\Delta^{+}$ \cite{Workman:2022ynf}. We wish to point out that the scale dependence of wave function overlap can be compensated for through the evaluation of constituent quark masses and hyperfine interaction terms from the experimental masses of the respective flavor sector. Therefore, we obtain the constituent quark masses and the hyperfine interaction terms in the strange and charm sectors from the experimental values of the strange and charm baryon masses, respectively, using (\ref{b1})-(\ref{b3}).
	
	In addition to isospin mass splitting in flavor $SU(2)$, we calculated the isospin mass splitting, \textit{i.e.}, $m_{d}-m_{u}$, for strange and charm sectors, respectively, as shown in Table \ref{isospin breaking}. We determine the isospin splitting as given in Column 2 of Table \ref{isospin breaking} by using the experimental masses of baryons  listed in Column 3 in Table \ref{isospin breaking} for the corresponding flavor sector. Since there are a limited number of known experimental masses for bottom baryons, we have restricted our calculations to the charm sector. It is interesting to note that the numerical values of the isospin mass splitting in flavors $SU(2)$ and $SU(3)$ are nearly the same and differ only in the second decimal place. However, the flavor mass splitting in $SU(4)$ is approximately $25\%$ smaller. Furthermore, our numerical value for isospin mass splitting for the charm sector is smaller when compared to the $2.494$ MeV obtained by Karliner \textit{et al.} \cite{Karliner:2019lau}. The difference in the numerical comparison can be attributed to different constituent quark masses in both works.
	
	%%%%%%%%%%%%%%%%%%%%%%%%%%%%%%%%%%%%%%%%%%%%%%%%%%%%%%%%%%%%%%%%%%
	\begin{table}[htp]
		\centering
		\captionof{table}{Isospin splitting between $u$ and $d$ quarks in different flavor sectors (in MeV).} 
		\label{isospin breaking}
		\begin{tabular}{|c|c|c|}	\hline 
			\textbf{Flavor Multiplet}  & \boldmath{$ m_{d} - m_{u}$ } & \textbf{Experimental Inputs \cite{Workman:2022ynf}} \\ \hline \hline
			Isospin sector & $ 2.957$ & $N, N^{*}$                       \\	
			Strange sector & $ 2.924$ & $\Sigma^{+}, \Sigma^{-}$         \\  
			Charm sector   & $ 2.190$ & $\Sigma_{c}^{0}, \Sigma_{c}^{+}$ \\	\hline 
		\end{tabular}
	\end{table}
	%%%%%%%%%%%%%%%%%%%%%%%%%%%%%%%%%%%%%%%%%%%%%%%%%%%%%%%%%%%%%%%%%%
	Proceeding further, we use numerical values given in Table \ref{Quark mass isospin} to obtain the masses of light as well as heavy (up to triply charm) baryons as shown in Tables \ref{mass_light} and \ref{mass_charm}, respectively \footnote{We follow the spectroscopic notation as per \cite{Workman:2022ynf} to list baryons in SU(3) multiplets.}. We compare our predictions to the experimental masses \cite{Workman:2022ynf} as well as other works \cite{Shah:2016mig, Shah:2016vmd, Shah:2017liu, Shah:2017jkr, Zhang:2009iya, Gutierrez-Guerrero:2019uwa}. We find that our predictions for masses are consistent with the current experimental data \cite{Workman:2022ynf}, with a maximum percentage error $\sim \mathcal{O}(3\%)$ with respect to experimental values. Our results for light baryons (including hyperons) are in good agreement with experimental values, where the percentage error is less than 2$\%$. In particular, for $J^P=\frac{3}{2}^+$ strange baryons, our predictions are in excellent agreement with experimental results. Our predictions for heavy flavor charm baryons compare well with experimental values, with a maximum percentage error of 3.26$\%$ (in the case of $\Sigma_c ^{(*)}$ baryons). It should be noted that the choice of $m_c \sim 1710$ MeV in our previous work \cite{Kumar:2005ei, Dhir:2009ax} provides a better agreement of singly charm baryon mass predictions with experimental values, where the hyperfine interaction terms $b_{uc},~b_{dc},~b_{sc},~ \text{and}~b_{cc}$ were calculated from symmetry relations. However, we have calculated $b_{uc},~b_{dc},~b_{sc},~\text{and}~b_{cc}$ interaction terms in the present work from current experimental data. Furthermore, for $m_c\sim 1710$ MeV, $b_{cc}$ acquires a large negative value, making it difficult to treat both singly and doubly heavy charm baryons on equal footing, as is the case in other theoretical models. We wish to emphasize that the choice of $m_{c}$ is crucial for determination of the hyperfine interaction term $b_{cc}$ appearing in doubly charm $J^P=\frac{1}{2}^+ ~\text{and}~\frac{3}{2}^+$ baryons, which is evident from the excellent agreement of our doubly charm baryon mass predictions with experiment and lattice QCD (LQCD) results \cite{Brown:2014ena}. As a result, we conclude that the current inputs of constituent quark masses and strong hyperfine interaction terms adequately explain the experimental data up to charm sector with a small margin of error.
	
	While comparing with other works, we find that the isospin symmetry is mostly kept intact; however, we provide comparisons over a range of models \cite{Shah:2016mig, Shah:2016vmd, Shah:2017liu, Shah:2017jkr, Zhang:2009iya, Gutierrez-Guerrero:2019uwa} that also include predictions of charged states. Most of the theoretical models use baryon masses as input for fitting of quark masses and model-dependent parameters, which are optimized to match the experimental masses. Because the hyperfine interaction terms are extracted from experimental masses, the EMS does not rely on any model-dependent parameters. Our mass predictions are in better agreement with experimental results as compared to QCD sum rules (QCDSR) \cite{Zhang:2009iya}, and contact interaction (CI) model \cite{Gutierrez-Guerrero:2019uwa}. The results from heavy baryon chiral perturbation theory (HB$\chi$PT) \cite{Jiang:2014ena} and hypercentral constituent quark model (hCQM) \cite{Shah:2016mig, Shah:2016vmd, Shah:2017liu, Shah:2017jkr} are in better agreement with the experiment than our results. These models, in contrast to our method, perform separate calculations guided by experimental results to predict the masses of singly and doubly heavy charm baryons and involve model-dependent parametrization. Furthermore, our prediction for the mass of doubly charm baryon, $M_{\Xi_{cc}^+}=$ 3623.81 MeV, is in excellent agreement with the latest measurement from LHCb \cite{LHCb:2021eaf}. In addition, compared to other investigations, our predictions in the doubly charm sector show good agreement with LQCD data.
	
	It is important to note that the Coleman-Glashow relation \cite{Coleman:1961jn} is reproduced from the isospin mass splitting obtained in our results, as follows:
	\begin{equation}
		\label{a1}
		\begin{gathered}
			M_{n} - M_{p}  = M_{\Sigma^{-}} - M_{\Sigma^{+}} + M_{\Xi^{0}} - M_{\Xi^{-}}.
		\end{gathered}
	\end{equation}
	The Coleman-Glashow relation (\ref{a1}) is well preserved despite of symmetry breaking effects because the strange quark mass and flavor $SU(3)$ hyperfine interaction terms, $b_{us}$, $b_{ds}$, and $b_{ss}$, appearing in $\Sigma$ and $\Xi$ baryons mass relations, for instance in (\ref{b2}), cancels out. As a result, the above relationship is governed by contributions from constituent quark masses $m_u$ and $m_d$, as well as strong hyperfine interaction terms between them ($b_{uu}$ and $b_{dd}$). Thus, sum rules among the masses and magnetic moments of baryons serve as a test for the group symmetry assumptions in theoretical models \cite{Yang:2020klp, Franklin:1975yu, Singh:1979js, Sharma:1981ua, Lipkin:1983cm}. The symmetry relations are inapplicable if the sum rules are not followed. In the Appendices \ref{appendixA} and \ref{appendixB}, we enlist several sum rules for masses and magnetic moments that are extended from light sector to heavy sector via quark transformations. 
	
	The experimental isospin mass splitting among the hadrons is important because it reflects true isospin contributions. We would like to emphasize that since our results are solely based on current experimental data, we expect our predictions for isospin mass splittings will be more reliable. After fixing the constituent quark masses and hyperfine interaction terms in Table \ref{Quark mass isospin}, the isospin mass splittings for uncharmed, singly charmed, and doubly charmed baryons are calculated, as shown in Tables \ref{isospin_strange}, \ref{isospin_charm}, and \ref{isospin_doubly_charm}, respectively. We compare our findings to experimental mass differences from PDG \cite{Workman:2022ynf} for the light (uncharmed) baryons. We observe that our predictions match well with the available experimental results. Further, in Table \ref{isospin_charm}, we compare our results for charmed baryons with those obtained from the pion mean-field approach \cite{Yang:2020klp}. We find that our results match well with the observations of the pion mean-field approach \cite{Yang:2020klp} for the account of the intrinsic mass difference between $u$ and $d$ quarks, and the strong hyperfine interactions. Our theoretical predictions exhibit differences with poorly known experimental PDG mean values. The average $M_{\Sigma_{c}^{0}} - M_{\Sigma_{c}^{++}}$ splitting of $-0.220(13)$\footnote{Values in the parentheses represent uncertainties.} MeV in PDG \cite{Workman:2022ynf} is well known discrepancy between theory and experiment. For $M_{\Sigma_{c}^{0}} - M_{\Sigma_{c}^{+}}$ and $M_{\Xi_{c}^{\prime0}} - M_{\Xi_{c}^{\prime+}}$, our predictions of $1.612$ MeV and $1.61$ MeV are close to the experimental values of $1.10(16)$ MeV and $0.8(6)$ MeV (within uncertainties), respectively. However, experimental results for $M_{\Xi_{c}^{0}} - M_{\Xi_{c}^{+}}=2.91(26)$ MeV and $M_{\Xi_{c}^{*0}} - M_{\Xi_{c}^{*+}}=0.85(58)$ MeV \cite{Workman:2022ynf}, are smaller than our estimates of $5.87$ MeV and $2.17$ MeV, respectively. We would like to point out that experimental mass splittings of $M_{\Sigma_{c}^{*0}} - M_{\Sigma_{c}^{*++}}$ have exceedingly large uncertainty, in fact greater than the central value \cite{Workman:2022ynf}, which is typical for multiple experimental observations. Unlike other theoretical approaches, we did not use the experimental information of isospin mass splitting of charmed baryons as input in our calculations owing to the poorly known measurements. 
	
	Furthermore, we predict the isospin mass splitting of the doubly charmed baryons $M_{\Xi_{cc}^{+}} - M_{\Xi_{cc}^{++}}$ to be $2.21$ MeV which is of comparable magnitude but opposite sign in comparison to theoretical models (Table \ref{isospin_doubly_charm}). We also predict the numerical value of the mass splitting, $M_{\Xi_{cc}^{*+}} - M_{\Xi_{cc}^{*++}}$, as $3.33$ MeV, which is larger in magnitude than $(-1.3^{+1.1}_{-1.2})$ MeV \cite{Wei:2015gsa}. It is worth noting that the aforementioned electromagnetic contributions to effective quark masses are expected to cancel out, and that strong hyperfine interaction contributions to effective quark masses result in better agreement with experimental mass splittings for strange baryons. Moreover, in our calculation, the sign of the splitting is governed by mass difference $m_d-m_u$, which is expected to be dominant contribution in heavy baryon sector. We anticipate experimental advances in this area in the coming years, and we hope that the growing interest in the heavy flavor sector will bring consensus between the theory and experiment. 
	
	\section {Screened Quark Charge}
	\label{SQC}
	According to EMS, the mass of quark is modified due to the interactions with neighboring quarks; similarly, the charge of a quark inside a baryon can also be affected. When a quark within a baryon is probed by a soft photon, the spectator quarks may shield the charge of the quark under scrutiny, thus altering its charge \cite{Dhir:2013nka, Dhir:2009ax, Kumar:2005ei, Verma:1986it}. This effect is entirely electromagnetic and is comparable to electron screening in atoms, where the presence of surrounding electrons in the inner shell causes a decrease in the effective nuclear charge on the valence electrons. The idea of screening is found not just in quantum electrodynamics with quark charge, but also in QCD as color charge screening. The effect of the shielding of quark charge varies with distance. The effective quark charge can be described as a linear function of the charges of the screening quarks. As a result, the effective charge of a quark $i$ inside a baryon $B(i~j~k)$ is calculated as follows:
	\begin{equation}    
		\label{e3}
		e_{i}^{\mathscr{E}}   =  e_{i}   +  \alpha_{i j} e_{j}   +  \alpha_{i k} e_{k}   ,
	\end{equation}
	where, $e_{i} $, $e_{j} $, and $e_{k} $ denote the bare charges of quarks \textit{i}, \textit{j}, and \textit{k}, respectively. In this case, the screening parameters of quark $i$ corresponding to the spectator quarks $j$ and $k$ are $\alpha_{ij}$ and $\alpha_{ik}$, respectively. Their determination will allow us to compute the effective charge and, as a result, the magnetic moments and other properties of baryons that result from the two-body interaction within baryons.
	
	In this study, we have assumed that the screened quark charge parameters are same across distinct independent flavor sectors. Consequently, we define them by choosing $\alpha_{i j} =\alpha_{j i}$ and invoking the isospin symmetry $(SU(2))$, as follows:
	\[\alpha_{uu}   =  \alpha_{ud}  =  \alpha_{dd} =  x ;\] 
	while, in strange sector $(SU(3))$:
	\[\alpha_{us}   =  \alpha_{ds}   =  \alpha_{ss  }   =  y;\] 
	and, for charm sector $(SU(4))$:
	\[\alpha_{uc}   =  \alpha_{dc}   =  \alpha_{sc}   =  \alpha_{cc}   =  z.\]
	It is worth noting that the effect of quark charge screening is expected to decrease with increasing mass, as heavier particles would have a point-like structure (effectively).
	
	Currently, precise experimental measurements of magnetic moments of all the octet baryons (except for $\Sigma^{0}$) and three of the decuplet baryons: $\Delta^{++}$, $\Delta^{+}$, and $\Omega^{-}$ are available \cite{Workman:2022ynf}. We utilize these magnetic moments as inputs and minimize with MINUIT \cite{James:1975dr} to determine the numerical values of the screening parameters, $x$ and $y$, \textit{i.e.}, 
	\ben \label{xy} x = 0.101(1);~ y = 0.136(2), \en
	which yields an excellent fit with $\chi^{2}$ value 0.073 (we have used (\ref{e4}) by replacing $X$ with magnetic moment, $\pmb{\mu}$, of the baryon). 
	The expression for theoretical magnetic moment operator, $\pmb{\mu}$, is given by
	\begin{equation}    
		\label{e5}
		\pmb{\mu } =\sum _{i} \frac{e_{i}   +  \alpha_{ij} e_{j}   +  \alpha_{ik}e_{k} }{2m_{i} ^{\mathscr{E}} } \pmb{\sigma _{i} }  =\sum _{i} \frac{e_{i}^{\mathscr{E}} }{2m_{i} ^{\mathscr{E}} }         \pmb{\sigma _{i} } ,  
	\end{equation}
	where, \textit{i = u, d, s}, and \textit{c}; $e_{i}^{\mathscr{E}}$ and $m_{i} ^{\mathscr{E}}$ represent the effective charge and effective mass of $i^{th}$ quark, respectively; and $\pmb{\sigma _{i}}$ denote the Pauli's spin matrices. Note that we have estimated the parameters $x$ and $y$ in \eqref{xy} without isospin symmetry breaking, where we used the values of constituent quark masses and hyperfine interaction terms, $b_{ij}$ from our recent work \cite{Hazra:2021lpa}.
	
	In the absence of experimental data on charmed baryon magnetic moments, the quark model and theoretical estimations from other models may be used to calculate the screened quark charge parameter $z$. It should be noted that Fomin \textit{et al.} \cite{Fomin:2019wuw} has utilized the measurements of higher multipole contributions to cascade radiative decays of $\psi(2s) \to \gamma_1 \chi_{c1,2}$ and $\chi_{c1,2}\to \gamma_2 J/\psi$. The normalized quadrupole contributions from the above mentioned decays are related to the anomalous magnetic moment and mass of the charm quark. The authors estimated the gyromagnetic factor for the charm quark ($g_c$) using the most accurate anomalous magnetic moment measurement from the recent BES III experiment \cite{BESIII:2017tsq}, which yields the magnetic moment for the $\mu_{\Lambda_c} \sim 0.48~\mu_N$ from $g_c/2m_c$. The result has limited precision due to uncertainties in charm quark mass and radiative corrections from strong interactions. On the other side, the magnetic dipole moment predictions of $\Lambda_c^+$ from various theoretical models, which range from ($0.34 - 0.43$) $\mu_N$ (with few exceptions), are likewise subject to charm quark mass uncertainty. Furthermore, Fomin \textit{et al.} \cite{Fomin:2019wuw} extend their analysis to fix the magnetic moments of $\Sigma_c^{++}$, $\Sigma_c^{+}$, $\Sigma_c^{0}$, and $\Xi_c^{+(0)}$ within the uncertainties introduced by the quark model. It should be noted that the results of Fomin \textit{et al.} \cite{Fomin:2019wuw} are extremely important since, as of today, the charm quark magnetic dipole moment is most precisely measured by quarkonium radiative decays \cite{Fomin:2019wuw, BESIII:2017tsq}. We use their results for magnetic moments of $\Lambda_c^{+}$, $\Sigma_c^{++}$, $\Sigma_c^{+}$, and $\Sigma_c^{0}$ to minimize screened charge parameter $z$ using (\ref{e5}), which is given by
	\ben \label{z} z = 0.023(1). \en
	The $\chi^2$ for the charm sector improves to $0.057$, as compared to the strange sector. Note that we use the same values of $x$ and $y$ from \eqref{xy} for the strange sector (being fixed from experimental magnetic moments). Additionally, we computed $z$ from each individual magnetic moments of $\Lambda_c^{+}$, $\Sigma_c^{++}$, $\Sigma_c^{+}$, and $\Sigma_c^{0}$ \cite{Fomin:2019wuw}, respectively, in order to assess the impact of uncertainties and determine an acceptable range of parameter $z$. These results are averaged to yield z = 0.155\footnote{$x$ and $y$ from \eqref{xy} are used as input.}. Considering the theoretical range ($0.34 - 0.43$) $\mu_N$, we use the lowest value of magnetic moment of $\Lambda_c^{+}$, $~0.34~\mu_N$, to minimize the $z$ parameter, which is surprisingly identical to \eqref{z}\footnote{$\chi^2$ is equal to 0.037.}. Similarly, we minimized $z$ for the highest value, $0.43~\mu_N$, to obtain $z = 0.013$ which correspond to a better $\chi^2=0.029$. Further, in contrast to our previous work \cite{Kumar:2005ei, Dhir:2009ax, Dhir:2013nka}, where $z$ parameter was estimated from strange sector using $SU(4)$ symmetry, we have now fixed three screened charge parameters, \textit{i.e.,} $~x$, $y$, and $z$, each one corresponding to isospin, strange, and charm sectors, respectively. 
	
	Finally, we now include the effects of isospin symmetry breaking on these parameters. We incorporate isospin broken effective quark masses in strange and charm sectors to obtain the magnetic moments. We anticipate that these parameter values will be changed in light of the revised effective quark masses obtained from isospin symmetry breaking analysis. Thus, the screened quark charge parameters are now fixed to 
	\ben \label{xyz}   x = 0.103(2); \
	y = 0.133(2); \
	z = 0.021(1), \en
	using the same methodology as described above for their determination. The $\chi^2$ for \eqref{xyz} are $0.068$ and $0.057$ for strange and charm sectors, respectively. It is evident that the screened charge parameters have quite small numerical variations. This can be understood from the trivial changes in the quark masses and hyperfine interaction terms that isospin symmetry breaking puts forth. As a result, we expect that the isospin breaking effects will be limited to the masses of baryons and will have minimal influence on magnetic moments. Furthermore, the soft photon used to detect magnetic moments can only observe the overall internal structure of the baryon. Thus, the magnetic moment (defined in terms of the quark magnetic moments by \eqref{e5}) contains a mass term, which is interpreted as the effective mass of the quark that appears in the denominator. We conclude, therefore, that changes in quark masses induced by isospin breaking will have a negligible effect on quark magnetic moments. In the next section, we compute the magnetic (transition) moments and M1 radiative decay widths of baryons, inclusive of effective quark masses, to see the effects of screened quark charge.
	
	\section{Magnetic Properties of Baryons}
	\label{properties}
	In this section, we calculate the magnetic moments of all the low lying ($l = 0$) baryon states up to the charm sector utilizing screened quark charge effects \eqref{e3} in addition to the EMS\footnote{Note that our EMS results do not include the effects of screened quark charge.}, henceforth referred to as the screened quark charge scheme (SQCS). Conventionally, the magnetic moments of baryons, $\mu_{B}$, can be obtained by inserting the magnetic moment operator, $\pmb{\mu }$, given in (\ref{e5}) between the appropriate baryon wave functions as given below.
	\begin{equation} \label{mm}  
		\mu_{B} = \langle\Psi_{sf}|\pmb{\mu }|\Psi_{sf}\rangle,
	\end{equation}
	where, $|\Psi_{sf}\rangle$ is the spin-flavor wave function of the corresponding baryon state. Further, we proceed to evaluate the transition moments ($\mu_{B^{\prime(*)}\to B^{(\prime)}}$) according to \eqref{mm} for $B^{\prime} \to B$ and $B^{*} \to B^{(\prime)}$. Following our earlier work \cite{Hazra:2021lpa}, we calculated transition moments using the geometric mean of the effective masses of the constituent quarks of the initial and final baryon states. Using the values from Table \ref{Quark mass isospin} and SQCS inputs given in \eqref{xyz}, we obtain the numerical results for the magnetic (transition) moments of the light and charm baryons in both EMS and SCQS, as shown in Tables \ref{octet_light_mm} - \ref{transition_charm}, respectively. In addition, the results corresponding to the $z$ values $0.021$ and $0.155$ for SQCS are given in columns 3 and 4, respectively, of Tables \ref{octet_charm_mm}, \ref{decuplet_charm_mm}, \ref{transition_charm}, and \ref{decay_width_charm}. Furthermore, we compare our results to other theoretical works and existing experimental results. Subsequently, we calculate the M1 radiative decay widths for the decay type $B^{\prime(*)}\to B^{(\prime)}\gamma$ using the following relation \cite{Dey:1994qi, Simonis:2018rld}: 
	\begin{equation}
		\label{e8}
		\Gamma_{B^{\prime(*)}\to B^{(\prime)}\gamma} = {\alpha\omega^3\over M^2_p}{2\over (2J+1)}|\mu_{B^{\prime(*)}\to B^{(\prime)}}|^{2},
	\end{equation}
	where, 
	\begin{equation}
		\label{e9}
		\omega = \frac{M^2_{B^{\prime(*)}} - M^2_{B^{(\prime)}}} {2M_{B^{\prime(*)}}},
	\end{equation}
	is the photon momentum in the rest frame of the decaying baryon. Here, $\alpha \approx~\frac{1}{137}$ is the fine structure constant, $M_p$ is the mass of proton, \textit{J} is the spin quantum number of the decaying baryon state, ${M_{B^{\prime(*)}}}$ and $M_{B^{(\prime)}}$ are the masses of initial and final baryon, respectively. The transition moments, $\mu_{B^{\prime(*)}\to B^{(\prime)}}$, are expressed in $\mu_N$. 
	
	As we pointed out in our previous work \cite{Hazra:2021lpa}, reliable predictions of M1 radiative decay widths need precise evaluation of photon momenta; to accomplish this, we relied upon experimentally available baryon masses and LQCD estimations \cite{Brown:2014ena}. Using $\omega$ values given in Tables XI and XII of our recent work \cite{Hazra:2021lpa}, we predict the decay widths for M1 radiative transitions as shown in Tables \ref{deacy_width_light} and \ref{decay_width_charm}. We also compare our results to other theoretical models.
	\section{Numerical Results and Discussions} \label{results}
	In this work, we predicted M1 radiative decay widths of strange and charmed baryons in SQCS using isospin-broken effective quark masses from EMS as well as screening of quark charge. We estimated isospin splitting, masses, magnetic, and transition moments up to charm baryons utilizing the inputs listed in Table \ref{Quark mass isospin} and \ref{xyz} during the evaluation process. The numerical results are given in Tables \ref{isospin breaking}, \ref{mass_light}, \ref{mass_charm}, \ref{isospin_strange}, \ref{isospin_charm}, \ref{isospin_doubly_charm}, \ref{octet_light_mm}, \ref{decuplet_light_mm}, \ref{octet_charm_mm}, \ref{decuplet_charm_mm}, \ref{transition_light}, \ref{transition_charm}, \ref{deacy_width_light}, and \ref{decay_width_charm}. In addition, we compare our results with experimental values and other theoretical models including HB$\chi$PT \cite{Jiang:2014ena, Sun:2014aya, Wang:2018gpl, Meng:2018gan, Wang:2018cre, Li:2017pxa}, hCQM \cite{Shah:2016mig, Shah:2016vmd, Shah:2017jkr, Shah:2017liu, Gandhi:2018lez}, QCDSR \cite{Zhang:2009iya}, CI \cite{Gutierrez-Guerrero:2019uwa}, LQCD \cite{Brown:2014ena, Borsanyi:2014jba, Can:2021ehb}, pion mean-field approach \cite{Yang:2020klp}, heavy quark symmetry (HQS) \cite{Hwang:2008dj}, chiral quark model ($\chi$QM) \cite{Linde:1997ni, Yu:2006sc}, chiral constituent quark model ($\chi$CQM) \cite{Sharma:2010vv}, bag model (BM) \cite{Simonis:2018rld}, NRQM \cite{Bernotas:2012nz}, relativistic three-quark model (RTQM) \cite{Faessler:2006ft}, covariant baryon chiral perturbation theory (B$\chi$PT) \cite{Shi:2018rhk, Liu:2018euh, Shi:2021kmm}, light cone QCD sum rule (LCQSR) \cite{Aliev:2015axa, Aliev:2001ig, Aliev:2008ay, Ozdem:2018uue, Aliev:2008sk, Ozdem:2019zis, Aliev:2004ju, Aliev:2009jt, Aliev:2014bma, Aliev:2016xvq}, hyper central model (HCM) \cite{Patel:2007gx, Patel:2008mv, Ghalenovi:2017fvw}, covariant spectator quark model (CSQM) \cite{Ramalho:2020tnn}, $\chi$PT \cite{Li:2017vmq}, chiral quark soliton model ($\chi$QSM) \cite{Yang:2019tst, Kim:2021xpp}, and constituent quark model (CQM) \cite{Wang:2017kfr}.
	\subsection{Masses and Magnetic Moments}
	As previously stated, we have included the isospin symmetry breaking results arising from the quark mass difference, $m_{d} - m_{u}$, in EMS. As described in Section \ref{isospin_splitting}, the implications of isospin symmetry breaking are smaller when compared to particle masses, which are known with high accuracy. We emphasize that our predictions for isospin mass splitting are reliable, since we only depend on the precise experimental data on baryon masses to compute the constituent quark masses and strong hyperfine interaction terms. Further, we find that our predictions for light and charm baryon masses agree well with existing experimental data, with a maximum percentage error of $\mathcal{O}(3\%)$ with regard to experimental values. In addition, we compare our results of masses and mass splittings for the light and charm baryon sectors to other theoretical works, as shown in Tables \ref{mass_light}, \ref{mass_charm}, \ref{isospin_strange}, \ref{isospin_charm}, and \ref{isospin_doubly_charm}. We observe that our results are in good agreement with experimental data and other theoretical approaches.
	
	Later, we extend our analysis to include the effect of screened quark charge in addition to EMS, \textit{i.e.}, in SQCS. Our results for magnetic moments of light baryon are shown in Tables \ref{octet_light_mm} and \ref{decuplet_light_mm} for both EMS and SQCS for the best fit given by \eqref{xyz}, which includes the effects of isospin symmetry breaking through masses. We found that the order of magnitude for isospin symmetry breaking effects is less than $2\%$ in light baryon magnetic moments, which is due to the relatively small mass difference in $u$ and $d$ quarks. As a result, we omit the distinction between isospin symmetry breaking and conserving cases from subsequent discussions \footnote{Moreover, our predictions for $z=0.013 ~\text{and}~ z=0.023$ in the isospin symmetry conserving case do not vary significantly from $z=0.021$ for charm baryons.}. It should be noted that that the results for strange baryons are only affected by the screening quark charge parameters $x$ and $y$. We predict the magnetic moments of charmed baryons for both schemes, as shown in Tables \ref{octet_charm_mm} and \ref{decuplet_charm_mm}. For the sake of comparison, we have considered an appropriate range for the screened quark charge parameter, $z$, in the charm sector, \textit{i.e.}, from (0.021 to 0.155). We also compare our results to those of other theoretical models for the charm baryons. We list our key findings as follows:
	
	\subsubsection{Magnetic Moments of $J^P=\frac{1}{2}^{(\prime)+}$ Baryons}
	\begin{enumerate}
		
		\item[i.] In Table \ref{octet_light_mm}, we compare our results for the magnetic moments of $J^P=\frac{1}{2}^+$ light baryons in SQCS to the experimental values \cite{Workman:2022ynf}. Intriguingly, when compared to EMS, the SQCS numerical values improve the compatibility with experimental numbers for the majority of the baryon magnetic moments. We would like to emphasize that screening of quark charge has an approximately $10\%$ effect on magnetic moments, with a few notable exceptions, such as $\Sigma^{-}$ and $\Xi^{-}$, which have a maximum change of $\mathcal{O}(27\%)$. Such deviations from EMS results can be attributed to quark charge screening. 
		
		\item[ii.] Due to the experimental difficulties associated with the short life time of $\Sigma^{0}$, its magnetic moment has not been measured yet; however, the magnetic moment of $\Sigma^{0}$ can be computed from the definition: $\mu_{\Sigma^{0}} = \frac{1}{2}(\mu_{\Sigma^{+}}^{Expt.} + \mu_{\Sigma^{-}}^{Expt.}) $ \cite{Barik:1986fp, Tsushima:2020gun}. The numerical estimate for $\Sigma^{0}$ in SQCS matches well with experimental expectations. 
		
		\item[iii.] For the singly charmed baryons, our results in SQCS present a good range corresponding to $z$ values given in columns 3 and 4 of Table \ref{octet_charm_mm}. We note that the magnetic moment results for $z=0.155$ exhibit a similar trend to those for $z=0.021$. Consequently, these results represent the widest range of magnetic moments in SQCS. Moreover, these numerical calculations over the extended range of $z$ compare favorably with the other theoretical model predictions \cite{Simonis:2018rld, Sharma:2010vv, Bernotas:2012nz, Faessler:2006ft, Shi:2018rhk}, indicating that our estimates for the screened quark charge parameter ($ z $) are credible. It is interesting to observe that our result for $\Lambda_c^{+}$ agrees well with the theoretical approaches \cite{Bernotas:2012nz, Faessler:2006ft, Sharma:2010vv}. Furthermore, LQCD \cite{Can:2021ehb} predicts $\mu_{\Sigma_c^{++}}= 2.220(505)~\mu_N$, $\mu_{\Sigma_c^{0}}=-1.073(269)~\mu_N$, and $\mu_{\Omega_c^{0}}= -0.639(88)~\mu_N$, which match well with our results for SQCS, within the uncertainties, except for $\mu_{\Omega_c^{0}}$. 
		
		\item [iv.] Our predictions for SQCS are compatible with improved BM \cite{Simonis:2018rld} results. The BM \cite{Simonis:2018rld} has employed independent scale factors corresponding to light and heavy quarks to imitate the mutual effects of center-of-mass motion, recoil, and other corrections. The marginal difference between BM \cite{Simonis:2018rld} and our results for the $\Xi_c^{(\prime)}$ baryons can be attributed to the state mixing effects that we are going to discuss later in Section \ref{state_mixing}. In contrast, we observe an average screening of $\mathcal{O}(6\%)$ in charmed baryons, with the exception of $\Sigma_c^{+}$, where the screening effect is maximum, \textit{i.e.}, $\mathcal{O}(24\%)$.
		
		\item[v.] It is interesting to note the consistency between theoretical models for $\Lambda-$ type singly heavy baryons, with a few exceptions. The numerical values of magnetic moments of $\Lambda-$ type singly heavy baryons, as noticed in \cite{Faessler:2006ft}, follow the leading contribution to the magnetic moment through the coupling of the photon to a heavy quark, resulting in agreement among different model predictions including ours. However, the contributions of the light quarks differ for the $\Sigma-$ type baryons, as observed in our results. Similarly, in $\chi$CQM \cite{Sharma:2010vv}, the magnetic moments of $\Lambda_c^+, ~\Xi_c^{+(0)}, ~\Omega_c^0$, and $\Omega_{cc}^+$ are dominated by valence contributions due to an excess of heavy quarks. 
		
		\item[vi.] We observe that the magnetic moment predictions in HB$\chi$PT \cite{Wang:2018gpl} are smaller than all the theoretical models including ours. In general, the dominant contribution in HB$\chi$PT is proportional to the leading term of $\mathcal{O}(1/m)$ and higher order corrections are of $\mathcal{O}(1/m\Lambda_{\chi}^2)$ that require estimation of model-dependent constants. The authors of HB$\chi$PT \cite{Wang:2018gpl} utilize LQCD results to obtain these constants, leading to smaller numerical predictions. Similarly, B$\chi$PT \cite{Shi:2018rhk} values that employ the extended-on-mass-shell (EOMS) technique to next-to-leading order are smaller than those of other models due to the low values of model parameters fitted using LQCD data.
		
		\item[vii.] We also compare our magnetic moment values with LCQSR \cite{Aliev:2015axa, Aliev:2001ig, Aliev:2008ay} predictions. The LCQSR approach utilizes the theoretical (QCD sum rules) and phenomenological correlation functions from dispersion relations to determine the magnetic moments of heavy baryons. The theoretical correlations are concerned with quarks and their interactions in the QCD vacuum, whereas the phenomenological correlation function is concerned with hadrons of the same flavor quantum numbers. We found that our results in SQCS match well with LCQSR \cite{Aliev:2015axa, Aliev:2001ig, Aliev:2008ay} predictions, with a few deviations.
		
		\item[viii.] For doubly charmed baryons, our predictions are consistent with all the other theoretical approaches \cite{Simonis:2018rld, Bernotas:2012nz, Faessler:2006ft}. However, the magnetic moment of $\Xi_{cc}^{++}$ differs from $\chi$CQM \cite{Sharma:2010vv}, which yields the lowest numerical value for magnetic moment among all theoretical models. Except for $\Xi_{cc}^{++}$, the LCQSR \cite{Ozdem:2018uue} estimates are around two times smaller than our results and predictions from other models. Their values, however, are consistent with LQCD \cite{Can:2021ehb} results, \textit{i.e.}, $\mu_{\Xi_{cc}^{+}}=0.425(29)~\mu_N$ and $\mu_{\Omega_{cc}^{+}}=0.413(24)~\mu_N$ and B$\chi$PT \cite{Liu:2018euh}. For SQCS magnetic moments of doubly charmed baryons, the screening has an effect of $\mathcal{O}(3\%)$. It is important to note that as we approach the doubly charmed baryons, the effect of screening reduces. 
	\end{enumerate}
	
	\subsubsection{Magnetic Moments of $J^P=\frac{3}{2}^+$ Baryons}
	\begin{enumerate}
		\item[i.] As observed for $J^P=\frac{1}{2}^+$ baryons, our numerical results are in very good agreement with experimental values (within errors) \cite{Workman:2022ynf} for the magnetic moments of $\Delta^{++}$, $\Delta^{+}$, and $\Omega^{-}$ when compared to EMS as well as other theoretical models \cite{Linde:1997ni, Sharma:2010vv, Ghalenovi:2017fvw}.
		
		\item[ii.]	For singly charmed $J^P=\frac{3}{2}^+$ baryons, our numerical predictions for SQCS are greater in magnitude than EMS. In contrast to EMS, our results in SQCS are more compatible with other models \cite{Simonis:2018rld, Sharma:2010vv, Bernotas:2012nz, Patel:2007gx, Patel:2008mv, Meng:2018gan}. As shown in Table \ref{decuplet_charm_mm}, HB$\chi$PT \cite{Meng:2018gan} results are in general smaller than other theoretical models. Moreover, apart from $\Sigma_c^{*0}$ and $\Xi_c^{*0}$, our SQCS results compare well with LCQSR \cite{Aliev:2008sk} values within the uncertainties.
		
		\item [iii.] Similarly, for the doubly charmed baryons, our results compare well with the predictions of $\chi$CQM \cite{Sharma:2010vv}, NRQM \cite{Bernotas:2012nz}, and HCM \cite{Patel:2007gx}, except for LCQSR \cite{Ozdem:2019zis} which predicts larger numerical values among other theoretical models. The numerical predictions for the magnetic moment of $\Xi_{cc}^{*+}$, however, differ between models. When we compare our result for $\Xi_{cc}^{*+}$ between EMS and SQCS, we observe a difference of an order of magnitude; the same can be seen in comparison with other theoretical predictions. The magnetic moment of $\Xi_{cc}^{*+}$ state presents a peculiar case with larger effect of screening. Again, the screening effect on magnetic moments is found to be significant, \textit{i.e.}, of $\mathcal{O} (15\%)$.
		
		\item[iv.] The magnetic moments of $\Xi_{cc}^{*+}$ and $\Omega_{cc}^{*+}$ are expected to be small as the contributions of the heavy quarks and light quark cancel out to some extent due to their opposite signs. However, in LCQSR \cite{Ozdem:2019zis}, the contribution from the light quark dominates over two charm quarks, with a large magnitude leading to larger numerical values.
		
		\item [v.] The triply heavy charmed baryon, $\Omega_{ccc}^{*++}$, presents an interesting case with three heavy quarks of the same flavor, and all the theoretical approaches \cite{Simonis:2018rld, Patel:2008mv, Bernotas:2012nz}, apart from $\chi$CQM \cite{Sharma:2010vv}, predict values that are broadly comparable. In addition, we observe a gradual reduction in the screening effect in doubly and triply heavy baryons.
	\end{enumerate}
	
	\subsection{Transition Moments}
	We have given our predictions for the light and charmed baryon transition moments for both EMS and SQCS in Tables \ref{transition_light} and \ref{transition_charm}. Likewise charm magnetic moments, we have listed our results for transition magnetic moments and M1 Decay widths at $z=0.155$. It is worth noting that our results for $z = 0.155$ quantify the maximum screening effect due to individual quark magnetic moments in transition magnetic moments and M1 decay widths. Further, due to different sign conventions, we only compare the magnitude of transition moments among different models. Following are our observations: 
	\begin{enumerate}
		\item[i.] In (${B^{\prime}\to B}$) transition moments for light baryons, our prediction $\mu_{\Sigma^{0}\rightarrow \Lambda^{0}}$ in SQCS is consistent with the experimental value within the error \cite{Workman:2022ynf}. It is interesting to note that $\Sigma^{0}\rightarrow \Lambda^{0}$ represents a peculiar case of a transition moment that is determined by the difference in magnetic moments of \textit{u} and \textit{d} quarks, with no plausible effect of isospin symmetry breaking. We re-emphasize that the majority of isospin breaking effects manifest in the masses, resulting in negligible isospin breaking in magnetic moments, and the difference in quark magnetic moments makes overall effects insignificant.
		
		\item[ii.] In contrast to magnetic moments, our results for transition moments in SQCS have smaller magnitudes as compared to EMS $\sim\mathcal{O} (10\%)$, which can be attributed to screening of quark charge. It should be highlighted that the magnitude of transition moments with regard to EMS results can be interpreted as an aggregation of effective quark charge (provided by \eqref{e3}) and magnetic moments of constituent quarks depending on their signs.
		
		\item[iii.] For (${B^{*}\to B^{(\prime)}}$), our predictions for light baryon transition moments for EMS are consistent with those of other models. Although our SQCS results for light baryon transition moments are smaller than other theoretical works \cite{Sharma:2010vv, Yu:2006sc, Ghalenovi:2017fvw, Ramalho:2020tnn,Li:2017vmq}, they are comparable to those of LCQSR \cite{Aliev:2004ju} predictions, with a few exceptions. In the absence of direct measurement, we compare our values to the $\Sigma^{*0}\rightarrow \Lambda^{0}$ and $\Sigma^{*+}\rightarrow \Sigma^{+}$ transition moment estimates extracted from \cite{CLAS:2011vzg, CLAS:2011iuw}, which are $20\%$ and $36\%$ larger, respectively. It can be seen that the transition moment predictions of $\chi$PT \cite{Li:2017vmq}, though consistent with experiment, are larger when compared to all the models. This is due to the fact that $\chi$PT \cite{Li:2017vmq} results are constrained by experimental observations from CLAS \cite{CLAS:2011vzg, CLAS:2011iuw}. Experimental observation of such transitions will guide the theoretical understanding of hadronic structure and properties.
		
		\item[iv.]	Our result for $\Sigma^{*-}\rightarrow \Sigma^{-}$ transition moment is smaller by a factor of two as compared to \cite{Sharma:2010vv, Yu:2006sc}. It should be noted that $\mu_{\Sigma^{*-} \rightarrow \Sigma^{-}}$ and $\mu_{\Xi^{*-} \rightarrow \Xi^{-}}$ are larger than EMS values. This is because, in addition to the increased effective charge of individual quarks due to screening, transition magnetic moment depends on the difference between the magnetic moments of $s$ and $d$ quarks, with a larger contribution from the latter. Furthermore, screening effects modify $\mu_{\Xi^{*-}\rightarrow \Xi^{-}}$ by a maximum of $\mathcal{O}(27\%)$, as compared to the average change of $12\%$.
		
		\item[v.] Similar to light baryons, our predictions for (${B^{\prime}\to B}$) charmed baryon transition magnetic moments in SQCS deviate from EMS results by $\sim\mathcal{O} (10\%)$. With a few exceptions, our results are generally consistent with existing theoretical models (as observed in magnetic moments of charm baryons). It is significant to note that for $\mu_{\Sigma_{c}^+\to \Lambda_{c}^+}$ and $\mu_{\Xi_{c}^{\prime +}\to \Xi_{c}^+}$, which have smaller transition moment values as compared to EMS, the magnetic moments of contributing quarks add constructively with a corresponding decrease in effective quark charge. We observe a similar pattern in results for $z=0.155$ in SQCS. The transition magnetic moment for $\Xi_{c}^{\prime 0} \rightarrow \Xi_{c}^{0}$ increases as the effective charge increases because the magnetic moments of $d$ and $s$ quarks add destructively, where the dominant contribution comes from the former. In addition, the effect of the enhanced screening parameter, $z=0.155$, is such that the contribution from $s$ quark magnetic moment becomes dominant in comparison to $d$, resulting in a reduction of the numerical value by nearly $20\%$.
		
		\item[vi.]	In general, our predictions in SQCS for the (${B^{*}\to B^{(\prime)}}$) charmed baryon transitions are consistent with BM \cite{Simonis:2018rld} and LCQSR \cite{Aliev:2009jt, Aliev:2014bma} estimates within errors, but are smaller than $\chi$CQM \cite{Sharma:2010vv}, and $\chi$QSM \cite{Yang:2019tst} models. Except for a few values, the numerical results of the HB$\chi$PT model \cite{Wang:2018cre} are the lowest when compared to all the other models, as seen in the case of magnetic moments. Their results for sextet to anti-triplet transition magnetic moments, $\Sigma_{c}^{*+} \rightarrow \Lambda_{c}^{+}$ and $\Xi_{c}^{*0} \rightarrow \Xi_{c}^{0}$, are consistent with our SQCS predictions except for $\Xi_{c}^{*+} \rightarrow \Xi_{c}^{+}$. These transition magnetic moments follow a mechanism similar to that described for octet to octet (${B^{\prime}\to B}$) charmed baryon transition moments. Therefore, we observe similar behaviour in SQCS for $z=0.155$.
		
		\item[vii.] It is interesting to note that, unlike our previous observations in the light and heavy baryon transition moments, the calculated values for sextet to sextet transition moments are greater than the EMS results. It should be emphasized that the reduced transition magnetic moments between neutral states caused by the increased screening charge parameter ($z=0.155$) highlights the significance of the underlying mechanism of individual quark magnetic moment contributions. In addition, the $\Sigma_{c}^{*+} \rightarrow \Sigma_{c}^{+}$ transition moment increases approximately by a factor of four when compared to EMS. This can be explained by the mutually dominant but destructive contributions of the magnetic moments of $u$ and $d$ quarks, which in turn become comparable to the magnetic moment contribution of the charm quark but with the opposite sign, resulting in a small total magnetic moment value. However, for enhanced screening quark charge parameter, the destructive contributions of $u$ and $d$ quark magnetic moments become dominant, resulting in an increased transition magnetic moment of $0.274~\mu_N$. In addition, the transition magnetic moment under investigation is sensitive to the selection of the involved parameters, which explains the wide range of numerical predictions in numerous approaches.
		
		\item[viii.] Further, $\Sigma_{c}^{*++} \rightarrow \Sigma_{c}^{++}$, $\Sigma_{c}^{*0} \rightarrow \Sigma_{c}^{0}$, and $\Omega_{c}^{*0} \rightarrow \Omega_{c}^{0}$ transition moments result from predominant contributions from light quark magnetic moments, as observed in various other theoretical models. Furthermore, $\mu_{\Xi_{c}^{*+}\to \Xi_{c}^{\prime +}} $ and $\mu_{\Xi_{c}^{*0}\to \Xi_{c}^{\prime 0}} $ can be interpreted similar to $\mu_{\Sigma_{c}^{*+} \rightarrow \Sigma_{c}^{+}}$ with destructive and constructive contributions of the magnetic moments of light quarks, respectively. In addition, we observe a change of $75\%$ and $20\%$ of their respective values at $z= 0.155$.
		
		\item[ix.] For doubly charm baryon transitions, our predictions are in good agreement with BM \cite{Simonis:2018rld} and $\chi$CQM \cite{Sharma:2010vv}; however, HB$\chi$PT \cite{Li:2017pxa} gives relatively larger predictions than other models. Also, transition magnetic moments involving doubly charm baryons receive dominant contributions from light quark magnetic moments. Overall, we observe a change of roughly $(25-35)\%$ as the screening charge parameter is enhanced.
	\end{enumerate}
	
	\subsection{Radiative Decay Widths}
	In this subsection, we present our predictions for the radiative M1 decay widths involving light and charmed baryons as shown in Tables \ref{deacy_width_light} and \ref{decay_width_charm}. As we have shown, the numerical value corresponding to $z=0.155$ can be interpreted as the maximum effect of quark charge screening based on the arguments established in the magnetic and transition moments. We shall therefore concentrate on the results of our best fit for quark charge screening. We have observed the following:
	\begin{enumerate}
		\item[i.] Since M1 transition decay width depends upon the magnitude of the transition moment and kinematic factors, the theoretical predictions are expected to follow a similar trend as observed in transition magnetic moments. Our predictions for light baryon M1 transitions in SQCS are consistent with those of $\chi$QM \cite{Yu:2006sc}. When compared to our results and $\chi$QM \cite{Yu:2006sc}, LCQSR \cite{Aliev:2004ju} predictions for M1 transitions are quite large, particularly for $\Delta \to N $, and $\Sigma^{*0}\rightarrow \Lambda^{0}$ transitions. This is in contrast to the results of transition moments for light baryons. It should be emphasised that most of the theoretical works underestimate the decay widths of $\Sigma^{*0}\rightarrow \Lambda^{0}$ and $\Sigma^{*+}\rightarrow \Sigma^{+}$ transitions, which were measured by the CLAS collaboration \textit{et al. }\cite{CLAS:2011vzg, CLAS:2011iuw} as $445(80)$ KeV and $250(70)$ KeV, respectively. However, the results of CSQM \cite{Ramalho:2020tnn} and $\chi$PT \cite{Li:2017vmq}, which are guided by experimental observations, are consistent. Such discrepancies in the light baryon sector have existed for a long time, indicating that additional experimental and theoretical efforts are required in such analyses. Furthermore, screening of quark charge has larger effect on the numerical predictions of radiative M1 transition on an average of $\mathcal{O}(23\%)$ in light baryon transitions, and as large as $\mathcal{O}(61\%)$ for $\Xi^{*-}\rightarrow \Xi^{-}$ transition. 
		
		\item[ii.] Our predictions for the singly charmed baryons are in very good agreement with those of BM \cite{Simonis:2018rld}, with the exception of the $\Xi_c^{\prime\,0}\rightarrow \Xi_c^{0}$ and $\Xi_c^{*0}\rightarrow \Xi_c^{0}$ transitions. As previously indicated, M1 decay widths in both SQCS and EMS predictions follow the same pattern as observed in transition magnetic moments. Apart from a few discrepancies, our results for $\frac{1}{2}^{\prime+} \to \frac{1}{2}^{+}$ M1 transition decay widths are more or less consistent when compared to other models. However, the LCQSR \cite{Aliev:2016xvq} predictions are smaller than our results. A similar pattern can be seen in the sextet to anti-triplet ($\frac{3}{2}^{+} \to \frac{1}{2}^{+}$) M1 decay widths, where most of the models agree to an order of magnitude for their predictions. On the other hand, CQM \cite{Wang:2017kfr} and $\chi$QSM \cite{Kim:2021xpp} predict decay widths that are roughly a factor of two larger and smaller than our values, respectively.
		
		\item[iii.] The $\chi$QSM \cite{Kim:2021xpp} predictions for (${B^{*}\to B^{(\prime)}}$) transitions, on the other hand, are smallest in magnitude among all theoretical models, with larger results for $\Sigma_c^{*+}\rightarrow \Sigma_c^{+}$ and $\Xi_c^{*+}\rightarrow \Xi_c^{\prime +}$. As discussed earlier, our results agree well with BM \cite{Simonis:2018rld}, though the theoretical estimate of HB$\chi$PT \cite{Wang:2018cre, Li:2017pxa} are generally small when compared to our results. The M1 decay widths of $\Sigma_c^{*+}\rightarrow \Sigma_c^{+}$ and $\Xi_c^{*+}\rightarrow \Xi_c^{\prime +} $ in various theoretical approaches ranges from $10^{-1}$ to $10^{-5}$ KeV, whereas our predictions in SQCS and EMS are of the order of $10^{-2}$ and $10^{-3}$ KeV, respectively. This is due to a mechanism similar to that described in transition magnetic moments. In addition, LCQSR \cite{Aliev:2009jt, Aliev:2014bma} and CQM \cite{Wang:2017kfr} predict larger decay widths when compared to our results for sextet to sextet ($\frac{3}{2}^{+}\rightarrow \frac{1}{2}^{\prime +}$) transitions. Moreover, the numerical values of transition moments and M1 decay widths for $\frac{3}{2}^{+}\rightarrow \frac{1}{2}^{+}$ radiative transitions are larger than those for $\frac{3}{2}^{+}\rightarrow \frac{1}{2}^{\prime+}$ due to the spin flip of the light quarks \cite{Kim:2021xpp}, which is true for almost all theoretical models, including ours. However, the $\Xi_c^{*0}\rightarrow \Xi_c^{0}$ transition remains an exception in this regard. Furthermore, our M1 decay widths predictions are nearly equal for $\Xi_c^{*0}\rightarrow \Xi_c^{\prime\,0}$ and $\Xi_c^{*0}\rightarrow \Xi_c^{0}$ transitions, whereas HB$\chi$PT \cite{Wang:2018cre} predicts large width for the latter. It is worth noting that the results of all theoretical models depend on different inputs and model-dependent parameters, which may lead to discrepancies among the results of different models.
		
		\item[iv.] In the case of radiative M1 decay widths of doubly charmed baryons, our SQCS results are fractionally smaller than those from EMS and BM \cite{Simonis:2018rld}; however, HB$\chi$PT \cite{Li:2017pxa} predictions are larger, roughly by an order of magnitude. The LQCD estimates \cite{Can:2021ehb} for $\Xi_{cc}^{*++}\to\Xi_{cc}^{++}$, $\Xi_{cc}^{*+}\to\Xi_{cc}^{+}$, and $\Omega_{cc}^{*++}\to\Omega_{cc}^{++}$ transition decay widths are $0.052(6)$ KeV, $0.065(4)$ KeV, and $0.056(1)$ KeV, respectively, which are exceedingly small compared to other models.
		
		\item [v.] The screening effect in the radiative decay widths of charmed baryons causes an average change of $20 \%$ in the numerical values, excluding the $\Sigma_c^{*+}\rightarrow \Sigma_c^{+}$ transition. In addition, corresponding to the maximum screening effect at $z=0.155$, the numerical results exhibit very large changes in the associated numerical values; however, these changes are still within acceptable ranges when compared to other theoretical models.
	\end{enumerate}
	
	The eventual availability of experimental numbers for radiative decay widths can help in testing the validity of the theoretical models. 
	\section{State mixing in flavor degenerate baryons}
	\label{state_mixing}
	The mixing of flavor degenerate baryons in strange and charm sectors, which contain three different quark flavors corresponding to different spin compositions, is of particular interest \cite{Franklin:1981rc}. We define, the physical states $|\mathscr{B} \rangle$ and $|\mathscr{B}^{\prime} \rangle$ as, 
	\begin{equation}
		\begin{gathered}
			\label{e11}
			|\mathscr{B} \rangle = \cos\theta |B \rangle  + \sin \theta |B^{\prime} \rangle , 	\\
			|\mathscr{B}^{\prime} \rangle = - \sin\theta|B \rangle + \cos \theta|B^{\prime} \rangle, 		
		\end{gathered}
	\end{equation}
	where, $|B \rangle$ (the first two quarks have relative spin $0$) and $|B^{\prime} \rangle$ (the first two quarks have relative spin $1$) are unmixed baryon states, and $\theta$ is the mixing angle. We use the mixing angle as given below \cite{Franklin:1981rc, Franklin:1996ve}, 
	\begin{equation}
		\begin{gathered}
			\label{e12}
			\tan{2\theta} = {\sqrt{3}\,(m_{j}-m_{i})\over2m_{k}-m_{i}-m_{j}},
		\end{gathered}
	\end{equation}  
	where, $m_{i}$, $m_{j}$, and $m_{k}$ are the constituent masses of quarks $i$, $j$, and $k$ inside the baryon, $B(ijk)$, respectively. We used isospin symmetry breaking induced constituent quark masses from Table \ref{Quark mass isospin} to calculate the mixing angle from \eqref{e12} for the strange and charm baryons. Furthermore, mixing in the magnetic moments of flavor degenerate baryons can be expressed as,
	\begin{equation}
		\begin{gathered}
			\label{e14}
			\mu(\mathscr{B}) = \mu_{B}\cos^2\theta + \mu_{B^{\prime}}\sin^2\theta + \mu_{B^{\prime}\to B}\sin2\theta ,\\
			\mu(\mathscr{B}^{\prime}) = \mu_{B}\sin^2\theta + \mu_{B^{\prime}}\cos^2\theta - \mu_{B^{\prime}\to B}\sin2\theta,
		\end{gathered}	
	\end{equation}
	where, $\mu_{B}$ and $\mu_{B^{\prime}}$ are the unmixed magnetic moments of $|B \rangle$ and $|B^{\prime} \rangle$ states that correspond to (\ref{e5}), respectively. The subsequent mixed transition moments are given by
	\begin{equation}
		\begin{gathered}
			\label{e15}
			\mu(\mathscr{B}^{\prime}\to \mathscr{B}) = \mu_{B^{\prime}\to B}\cos2\theta - {1\over2}(\mu_{B} - \mu_{B^{\prime}})\sin2\theta ,\\
			\mu(\mathscr{B}^{*}\to \mathscr{B}^{(\prime)}) = \mu_{B^{*}\to B^{(\prime)}}\cos\theta \pm \mu_{B^{*}\to B^{\prime ()}}\sin\theta, 
		\end{gathered}	
	\end{equation}
	where, $\mu_{B^{\prime(*)}\to B^{(\prime)}}$ represents the unmixed transition moments of baryons.
	
	The intent of state mixing is to compensate for the ordering of the quarks. The preferred ordering of quarks is such that the least amount of mixing is achieved. This is accomplished by arranging the quarks in flavor degenerate baryons in increasing order of their masses to obtain the smallest $(m_j-m_i)$, which results in minimal mixing \cite{Franklin:1981rc, Simonis:2018rld}. Further, the mixing effects propagate to the decay widths via mixed transition moments.
	\subsection{Effect of State Mixing}
	The physical significance of the mixed state lies in the fact that it represents a physically observed state in experiments. The mixing of baryon masses is neglected, as it is of second order in the mixing angle. On the other hand, the effect of mixing is much more visible in magnetic (transition) moments, being of the first order, which in turn affects the M1 radiative decay widths substantially \cite{Franklin:1981rc}. Thus, the numerical results of magnetic (transition) moments and the radiative decay widths obtained by considering the state mixing are listed in Tables \ref{mixing_mm} and \ref{mixing_decay_width}, along with results from other models \cite{Simonis:2018rld, Simonis:2018rld, Bernotas:2012nz}. We also list our unmixed results in order to provide a thorough comparison. Our observations are listed as follows:
	
	\begin{enumerate}
		\item[i.] The numerical values of mixing angle increase with baryon masses (except for \textit{udc} states), which results from the mass difference between the first two quarks, as shown in the Table \ref{mixing_mm}. The mixing angle associated with the \textit{udc} states is negligible because of the smaller value of the numerator in \eqref{e12}. Our $\Sigma- \Lambda$ type mixing results match well with NRQM \cite{Bernotas:2012nz} and BM \cite{Simonis:2018rld}. For \textit{usc} states, our numerical estimates for mixing between $\Xi_c-\Xi_c^{\prime}$ type are consistent when compared to $3.8^{\circ}$ in QM \cite{Franklin:1981rc, Franklin:1996ve} and $5.5(1.8)^{\circ}$ in QCDSR \cite{Aliev:2010ra}. In addition, we observe that the effect of state mixing in $uds$ and $udc$ states is negligibly small, ranging up to 3$\%$. However, the mixing effect in $\Sigma_{c}^{*+} \rightarrow \Sigma_{c}^{+}$ is $\mathcal{O}(10\%)$.
		
		\item[ii.] As expected, mixing has significant effects of $\mathcal{O}(50\%)$, on the magnetic properties of $usc$ and $dsc$ states. We observe large mixing in $\Xi_{c}^{*+}\rightarrow\Xi_{c}^{\prime\,+}$, $\Xi_{c}^{\prime\,0}\rightarrow\Xi_{c}^{0}$, and $\Xi_{c}^{*0}\rightarrow\Xi_{c}^{0}$ transitions, which contributes to their transition moments and decay width results, as shown in Table \ref{mixing_decay_width}. Furthermore, our predictions are consistent with other theoretical models except for $\Xi_{c}^{\prime\,0}\rightarrow\Xi_{c}^{0}$ in BM \cite{Simonis:2018rld}. Apart from this, mixing has substantial effect on the magnetic moments of $\Xi_{c}^{+}$ and $\Xi_{c}^{\prime\,+}$ states resulting in a respective change of $48\%$ and $29\%$ in their values.
	\end{enumerate}
	
	\section{Summary and Conclusions}
	\label{conclusions}
	In the present work, we primarily focus on the screening of quark charge in SQCS, which incorporates isospin symmetry breaking through effective quark masses. We made predictions of the magnetic properties, \textit{i.e.}, magnetic (transition) moments and M1 radiative decay widths, of all the ground state baryons up to triply charm. We used precise experimental information in the isospin sector to obtain constituent \textit{u} and \textit{d} quark masses and their hyperfine interaction terms ($b_{uu},~ b_{ud},~\text{and}~b_{dd}$) through minimization to improve upon our previous results in EMS. In a similar fashion, we have estimated isospin splitting in strange and charm baryons in their respective flavor sectors. We have calculated the constituent quark masses and hyperfine interaction terms from the precise experimental data in a model-independent way, individually, for the strange and charm flavors. Next, we calculated the masses of low lying ($J^P=\frac{1}{2}^+\text{and}~\frac{3}{2}^+$) baryon states up to $C=3$. We then used available experimental magnetic moments to evaluate the screened quark charge parameters ($\alpha_{ij}$) for light baryons. After evaluating all the ingredients, we made robust predictions for magnetic (transition) moments and M1 decay widths for light baryons up to the strange flavor. Proceeding in a similar way, we extended our analysis to observe charmed baryon masses and magnetic properties. In the absence of experimental magnetic moments, we rely on the analysis by Fomin \textit{et al.} \cite{Fomin:2019wuw}, which explores experimental prospects of magnetic moment of charm baryons, to determine the screened quark charge parameter. Then, we predicted the magnetic (transition) moments and decay widths of charmed baryons in both SQCS and EMS. In addition, we include the mixing of flavor degenerate baryon states in our analysis.
	As mentioned earlier, the variations introduced by quark charge screening and state mixing effects are carefully observed and analyzed through the electromagnetic properties of baryons. In light of this, we arrived at the following conclusions:
	\begin{itemize}
		\item We believe that because our numerical values for constituent quark masses and hyperfine interaction terms for light and charm flavor sectors are evaluated from experimental data, they are more reliable. Consequently, our predictions for the masses of light and charm baryons are in good agreement with experimental values, with a maximum percentage error $\sim \mathcal{O}(3\%)$ in the case of $\Sigma_c ^{(*)}$ baryons. Unlike various theoretical models, we treated singly and doubly charm baryons on a same footing in a model-independent analysis.
		
		\item Our predictions of the isospin mass splitting for strange baryons are in very good agreement with the available experimental data. However, a comparison of theoretical and experimental splitting in charm baryon data reveals inconsistencies because of poor experimental values with significant uncertainties.
		\item We predict the isospin mass splitting of the doubly charmed baryons as $M_{\Xi_{cc}^{+}} - M_{\Xi_{cc}^{++}}=2.21$ MeV and $M_{\Xi_{cc}^{*+}} - M_{\Xi_{cc}^{*++}}=3.33$ MeV, which is yet to be observed experimentally. We wish to point out that our prediction for mass of $\Xi_{cc}^{+}$ is in excellent agreement with recent experimental result from LHCb \cite{LHCb:2021eaf}. Furthermore, we observed nearly uniform isospin splitting in strange and charm baryons, \textit{i.e.}, around $(2 - 3)$ MeV.
		\item Numerically, magnitude of isospin mass splitting is very small as compared to masses of baryons, which are precisely measured. Therefore, isospin symmetry breaking will have negligible effect on magnetic moments, where the mass of quark appear in the denominator.
		\item Our analysis shows that the quark charge screening has a significant effect on the magnetic moments of strange baryons, with an overall average of $17\%$. Furthermore, the inclusion of quark charge screening improved consistency of magnetic and transition moments predictions with the strange baryon experimental observations, with few exceptions.
		\item Interestingly, the effect of screening gradually decreases in magnetic moments from light to heavy flavor owing to the variation of the screened charge parameter, when the size of the baryon is expected to decrease. In singly charmed baryons in magnetic moments, we find an average screening effect of $\mathcal{O}(6\%)$, with an exceptional screening effect of $24\%$ in the case of $\Sigma_c^+$, and it reduces to $3\%$ in the case of doubly charm baryons. 
		\item Although magnetic transition moments and decay widths show a similar trend of declining screening effect from light to heavy sector, the effect is more prominent. We find that increasing or decreasing magnitudes of numerical values of transition magnetic moments (M1 decay widths) can be explained as a consequence of the accumulation of individual magnetic moments of constituent quarks with respective signs and effective quark charge due to screening. The magnetic moments of constituent quarks add constructively or destructively along with the effective charge of quarks, which can produce a screening effect of $\mathcal{O}(10\%)$ or more.
		\item We find that, despite appearing to be consistent, numerical predictions for transition magnetic moments and, consequently, M1 decay widths across various theoretical approaches provide some interesting results, such as $\Sigma_c^{*+}\rightarrow \Sigma_c^{+}$. Experimental measurement of such transitions can shed some light on the internal structure of the heavy baryons.
		
		\item We have focused on the effect of state mixing in magnetic moments rather than in masses due to the first order dependence of magnetic moments on mixing angle. We found that the mixing is larger in heavier baryon states, except for \textit{udc}. In our analysis for the mixing of states, we deduce that the mixing effects, though relatively small, have improved our results in comparison with the available experimental numbers in the strange baryons. However, in the charm baryons, our results showed considerably large effects of mixing in magnetic (transition) moments. Therefore, we conclude that the effects of mixing are of much importance in M1 radiative decay widths (since, $ \Gamma_{B^{\prime(*)}\to B^{(\prime)}\gamma} \propto |\mu_{B^{\prime(*)}\to B^{(\prime)}}|^{2}$) with few exceptions.
	\end{itemize}
	We hope that our analysis will be useful for upcoming theoretical and experimental studies on the structure and properties of heavy flavor baryons.

	\section*{Acknowledgment}
	The authors gratefully acknowledge the financial support by the Department of Science and Technology (SERB:CRG/2018/002796), New Delhi.

	%%%%%%%%%%%%%%%%%%%%%%%%%%%%%%%%%%%%%%%%%%%%%%%%%%%%%%%%%%%%%%%%%%%%%%%%%%%%%%%%%%%%%%%%
	\newpage
	\appendix
	
	\section*{Appendix}
	%%%%%%%%%%%%%%%%%%%%%%%%%%%%%%%%%%%%%%%%%%%%%%%%%%%%%%%%%%%%%%%%%%%%%%%%%% 
	\section{Effective Quark Masses}
	\label{bar_mass_exp}
	In effective mass scheme, the mass of the quark inside a baryon gets modified by the interaction with other two quarks through one-gluon exchange interaction. We have listed the octet and decuplet baryon mass expressions as a sum of constituent quark masses and hyperfine interaction terms \cite{Hazra:2021lpa} as follows:
	
	For $(iik)-$type $J^P = \frac{1}{2}^+$ baryons, we can write
	\begin{equation}
		\begin{gathered}
			\label{A1}
			m_i^\mathscr{E} = m_j^\mathscr{E} = m + \alpha{b}_{ij} + \beta{b}_{ik} ,\\
			m_k^\mathscr{E} = m_k + 2\beta{b}_{ik} ,\\
		\end{gathered}
	\end{equation}
	where, $m_i = m_j = m$ and $b_{ik} = b_{jk}$. Throughout the discussions each $\it{i,~j,~k}$ represents $\it{u, d, s,}$ and $\it{c}$ quarks. The $\alpha$ and $\beta$ parameters are calculated from $\bf{s_i.s_j}$ as follows:
	\[ {M}_{B} = 2m + m_k + \frac{b_{ij}}{4} - b_{ik},\]
	from
	\[	{\bf{s_i.s_j}} = \frac{1}{4},~{\bf{{s_i.s_k} = {s_j.s_k}}} = -\frac{1}{2} , \]
	we get, 
	\[ \alpha = \frac{1}{8}\mbox{ and }\beta = -\frac{1}{4}.\]
	Further, we can generalize equation (\ref{e1}) for $J^P = \frac{1}{2}^+$ baryons as
	\[  M_{{B}_{\frac{1}{2}^+}} = m_i + m_j + m_k + \frac{b_{ij}}{4} - \frac{b_{jk}}{2} - \frac{b_{ik}}{2}, \]
	leading to 
	\begin{equation}
		m_i^\mathscr{E} =  m_j^\mathscr{E} = m + \frac{b_{ij}}{8} - \frac{b_{ik}}{4},
	\end{equation}
	\text{and}
	\begin{equation}
		m_k^\mathscr{E}= m_k - \frac{b_{ik}}{2};\mbox{ for } i = j\ne k.
	\end{equation}
	In addition, we can further give the effective mass expressions for $\Sigma-$ and $\Lambda-$ type, defined in \eqref{b1} and \eqref{b2}, and $J^P = \frac{3}{2}^+$ baryons (as in \eqref{b3}) from equation (\ref{A1}), which are given by
	\begin{itemize}
		\item [I.] For $(ijk)$ $\Lambda-$type,
		\begin{equation}
			\begin{gathered}
				m_i^\mathscr{E} = m_i - \frac{3b_{ij}}{8},\\
				m_j^\mathscr{E} = m_j - \frac{3b_{ij}}{8},
			\end{gathered}
		\end{equation}
		and
		\begin{equation}
			m_k^\mathscr{E} = m_k; \mbox{ for } i\ne j\ne k.
		\end{equation}
		
		\item[II.] For $(ijk)$ $\Sigma-$type,
		\begin{equation}
			\begin{gathered}
				m_i^\mathscr{E} = m_i + \frac{b_{ij}}{8} - \frac{b_{ik}}{4},\\
				m_j^\mathscr{E} = m_j + \frac{b_{ij}}{8} - \frac{b_{jk}}{4},\\
			\end{gathered}
		\end{equation}
		and 
		\begin{equation}
			m_k^\mathscr{E} = m_k - \frac{b_{jk}}{4} - \frac{b_{ik}}{4}; \mbox{ for } i\ne j\ne k.
		\end{equation}
		
		\item [III.] For $J^P = \frac{3}{2}^+$ baryons,
		\[  M_{{B}_{\frac{3}{2}^+}} = m_i + m_j + m_k + \frac{b_{ij}}{4} + \frac{b_{jk}}{4} + \frac{b_{ik}}{4}, \]
		we get \[ \alpha = \beta = \frac{1}{8}. \]
		The effective quark masses corresponding to quark order are given by
		\begin{itemize}
			\item [a)] For $(iik)-$type,
			\begin{equation}
				\label{a8}
				m_i^\mathscr{E} =  m_j^\mathscr{E} = m + \frac{b_{ij}}{8} + \frac{b_{ik}}{8},
			\end{equation}
			and
			\begin{equation}
					\label{a9}
				m_k^\mathscr{E}= m_k + \frac{b_{ik}}{4}; \mbox{ for } i = j\ne k.
			\end{equation}
			
			\item [b)] For $(ijk)-$type,
			\begin{equation}
				\begin{gathered}
					m_i^\mathscr{E} = m_i + \frac{b_{ij}}{8} + \frac{b_{ik}}{8},\\
					m_j^\mathscr{E} = m_j + \frac{b_{jk}}{8} + \frac{b_{ij}}{8},\\
				\end{gathered}
			\end{equation}
			and
			\begin{equation}
				m_k^\mathscr{E} = m_k + \frac{b_{ik}}{8} + \frac{b_{jk}}{8}; \mbox{ for } i\ne j\ne k.
			\end{equation}
			
			\item [c)] For $(iii)-$type,
			\begin{equation}
					\label{a12}
				m_i^\mathscr{E} = m_j^\mathscr{E} = m_k^\mathscr{E} = m + \frac{b_{ij}}{4},
			\end{equation}
			and
			\[	b_{ij} = b_{jk} = b_{ik}; \mbox{ for } i = j = k.\]
		\end{itemize}
	\end{itemize}
	
	Using the above given set of equations, we have calculated the constituent quark masses and strong hyperfine interaction terms, $b_{ij}$, corresponding to inputs as shown in Table \ref{Quark mass isospin}. For example, as mentioned in the section \ref{isospin_splitting}, we have obtained $m_u$, $m_d$, $b_{uu}$, and $b_{dd}$ by using $N~\text{and}~N^*$  experimental masses as inputs, as shown in Table \ref{Quark mass isospin}. We minimize $(i~i~k)-$type and $(i~i~i)-$type mass relations for $\frac{1}{2}^+$ and $\frac{3}{2}^+$ baryons masses, using \eqref{a2}, \eqref{a3}, \eqref{a8}, \eqref{a9}, and \eqref{a12}. Similarly, the corresponding mass relations described from (A1)-(A12) can be used to extract the numerical values listed in Columns 2 and 4 of Table \ref{Quark mass isospin}.     
	\section{Mass Sum Rules}
	\label{appendixA}
	In the following we are listing the mass sum rules for charmed as well as uncharmed baryons \cite{Yang:2020klp, Franklin:1975yu, Singh:1979js, Sharma:1981ua, Lipkin:1983cm}.
	\begin{enumerate}
		\item $ M_{n} - M_{p} = M_{\Delta^{0}} - M_{\Delta^{+}}  \approx 1.290 MeV $
		\item $\displaystyle \underbrace{ M_{n} - M_{p}} _\text{1.294 MeV} =  \underbrace{M_{\Sigma^{-}} -  M_{\Sigma^{+}} + M_{\Xi^{0}} - M_{\Xi^{-}}}_\text{1.290 MeV} = \underbrace{ M_{\Sigma^{*-}} -  M_{\Sigma^{*+}} + M_{\Xi^{*0}} - M_{\Xi^{*-}}}_\text{1.290 MeV}$
		\begin{enumerate}
			\item $s \to c$
			\newline $\displaystyle \underbrace{ M_{n} - M_{p}} _\text{1.294 MeV}  = \underbrace{M_{\Sigma_{c}^{0}} -  M_{\Sigma_{c}^{++}} + M_{\Xi_{cc}^{++}} - M_{\Xi_{cc}^{+}}}_\text{1.300 MeV} =\underbrace{ M_{\Sigma_{c}^{*0}} -  M_{\Sigma_{c}^{*++}} + M_{\Xi_{cc}^{*++}} - M_{\Xi_{cc}^{*+}}} _\text{1.300 MeV}$ 
			\item $u \to s$, $s \to c$
			\newline $\displaystyle \underbrace{ M_{\Sigma^{-}} - M_{\Xi^{-}}} _\text{-145.5 MeV}  = \underbrace{M_{\Sigma_{c}^{0}} -  M_{\Omega_{c}^{0}} + M_{\Omega_{cc}^{++}} - M_{\Xi_{cc}^{+}}}_\text{-145.5 MeV} =\underbrace{ M_{\Sigma_{c}^{*0}} -  M_{\Omega_{c}^{*0}} + M_{\Omega_{cc}^{*++}} - M_{\Xi_{cc}^{*+}}} _\text{-145.5 MeV}$
			\item $d \to s$, $s \to c$
			\newline $\displaystyle \underbrace{ M_{\Xi^{0}} - M_{\Sigma^{+}}} _\text{146.8 MeV}  = \underbrace{M_{\Omega_{c}^{0}} -  M_{\Sigma_{c}^{++}} + M_{\Xi_{cc}^{++}} - M_{\Omega_{cc}^{+}}}_\text{146.8 MeV} =\underbrace{M_{\Omega_{c}^{*0}} -  M_{\Sigma_{c}^{*++}} + M_{\Xi_{cc}^{*++}} - M_{\Omega_{cc}^{*+}}}_\text{146.8 MeV}$
		\end{enumerate}   
		\item $\displaystyle \underbrace{ M_{\Sigma^{+}} + M_{\Sigma^{-}} - 2M_{\Sigma^{0}}}_\text{-0.260 MeV} = \underbrace{M_{\Delta^{++}} +  M_{\Delta^{0}}- 2M_{\Delta^{+}}}_\text{-0.280 MeV} = \underbrace{M_{\Sigma^{*+}} + M_{\Sigma^{*-}} - 2M_{\Sigma^{*0}}}_\text{-0.280 MeV} $
		\begin{enumerate}
			\item $s \to c$
			\newline $\displaystyle \underbrace{M_{\Sigma_{c}^{++}} + M_{\Sigma_{c}^{0}} - 2M_{\Sigma_{c}^{+}}}_\text{-0.270 MeV} = \underbrace{M_{\Delta^{++}} +  M_{\Delta^{0}}- 2M_{\Delta^{+}}}_\text{-0.280 MeV} = \underbrace{M_{\Sigma_{c}^{*++}} + M_{\Sigma_{c}^{*0}} - 2M_{\Sigma_{c}^{*+}}} _\text{-0.270 MeV} $ 
			\item $u \to s$, $s \to c$
			\newline $\displaystyle \underbrace{M_{\Omega_{c}^{0}} + M_{\Sigma_{c}^{0}} - 2M_{\Xi_{c}^{\prime 0}}}_\text{-3.640 MeV} = \underbrace{M_{\Omega^{-}} + M_{\Sigma^{*-}} - 2M_{\Xi^{*-}}}_\text{-3.630 MeV} = \underbrace{M_{\Omega_{c}^{*0}} + M_{\Sigma_{c}^{*0}} - 2M_{\Xi_{c}^{* 0}}}_\text{-3.640 MeV}$
			\item $d \to s$, $s \to c$
			\newline $\displaystyle \underbrace{M_{\Sigma_{c}^{++}} + M_{\Omega_{c}^{0}} - 2M_{\Xi_{c}^{\prime +}}}_\text{-3.930 MeV} = \underbrace{M_{\Delta^{++}} + M_{\Xi^{*0}} - 2M_{\Sigma^{*+}}}_\text{-3.930 MeV} = \underbrace{M_{\Sigma_{c}^{*++}} + M_{\Omega_{c}^{*0}} - 2M_{\Xi_{c}^{* +}}}_\text{-3.930 MeV}$
		\end{enumerate} 
		\item $\displaystyle \underbrace{M_{\Xi^{0}} - M_{\Sigma^{+}}}_\text{146.8 MeV} = \underbrace{\frac{1}{3}({M_{\Omega^{-}} -  M_{\Delta^{++}}})}_\text{146.8 MeV} = \underbrace{M_{\Xi^{*0}} - M_{\Sigma^{*+}}}_\text{146.8 MeV}$
		\begin{enumerate}
			\item $s \to c$
			\newline $\displaystyle \underbrace{M_{\Xi_{cc}^{++}} - M_{\Sigma_{c}^{++}}}_\text{1248 MeV} = \underbrace{\frac{1}{3}({M_{\Omega_{ccc}^{*++}} -  M_{\Delta^{++}}})}_\text{1248 MeV} = \underbrace{M_{\Xi_{cc}^{*++}} - M_{\Sigma_{c}^{*++}}}_\text{1248 MeV}$
			\item $u \to s$, $s \to c$
			\newline $\displaystyle \underbrace{ M_{\Omega_{cc}^{+}} -  M_{\Omega_{c}^{0}}}_\text{1101 MeV} = \underbrace{\frac{1}{3}({M_{\Omega_{ccc}^{*++}} -  M_{\Omega^{-}}})}_\text{1101 MeV} = \underbrace{ M_{\Omega_{cc}^{*+}} -  M_{\Omega_{c}^{*0}}}_\text{1101 MeV}$
		\end{enumerate} 
		\item $\displaystyle \underbrace{M_{\Sigma^{*+}} - M_{\Sigma^{*-}} + 2 (M_{\Xi^{*-}} - M_{\Xi^{*0}})}_\text{-0.280 MeV} = \underbrace{M_{\Sigma_{c}^{*++}} - M_{\Sigma_{c}^{*0}} + 2 (M_{\Xi_{c}^{*0}} - M_{\Xi_{c}^{*+}})}_\text{-0.290 MeV} $    
	\end{enumerate}   
	%%%%%%%%%%%%%%%%%%%%%%%%%%%%%%%%%%%%%%%%%%%%%%%%%%%%%%%%%%%%%%%%%%%%%%%%%% 	
	\section{Magnetic Moments Sum Rules}
	\label{appendixB}
	The sum rules for magnetic moments of charmed and uncharmed baryons are listed as \cite{Yang:2020klp, Franklin:1975yu, Singh:1979js, Sharma:1981ua, Lipkin:1983cm},
	\begin{enumerate}
		\item $\displaystyle \underbrace{\mu_{p} - \mu_{n}}_\text{4.661 $\mu_N$} = \underbrace{\mu_{\Sigma^{+}} - \mu_{\Sigma^{-}} + \mu_{\Xi^{-}} - \mu_{\Xi^{0}}}_\text{4.507 $\mu_N$ }$ [Coleman - Glashow relation]
		\begin{enumerate}
			\item $s \to c$
			\newline $\displaystyle \underbrace{\mu_{p} - \mu_{n}}_\text{4.661 $\mu_N$} = \underbrace{\mu_{\Sigma_{c}^{++}} - \mu_{\Sigma_{c}^{0}} + \mu_{\Xi_{cc}^{+}} - \mu_{\Xi_{cc}^{++}}}_\text{4.568 $\mu_N$ } \text{[Coleman - Glashow relation in charm sector]}$ 
			\item $u \to s$, $s \to c$
			\newline $\displaystyle \underbrace{\mu_{\Xi^{-}} - \mu_{\Sigma^{-}}}_\text{0.602 $\mu_N$} = \underbrace{\mu_{\Omega_{c}^{0}} - \mu_{\Sigma_{c}^{0}} + \mu_{\Xi_{cc}^{+}} - \mu_{\Omega_{cc}^{+}}}_\text{0.420 $\mu_N$ }$
			\item $d \to s$, $s \to c$
			\newline $\displaystyle \underbrace{\mu_{\Sigma^{+}} - \mu_{\Xi^{0}}}_\text{3.906 $\mu_N$} = \underbrace{\mu_{\Sigma_{c}^{++}} - \mu_{\Omega_{c}^{0}} + \mu_{\Omega_{cc}^{+}} - \mu_{\Xi_{cc}^{++}}}_\text{4.148 $\mu_N$ }$
		\end{enumerate}
		\item $\displaystyle \underbrace{3 (\mu_{p} + \mu_{n})}_\text{3.060 $\mu_N$} = \underbrace{\mu_{\Sigma^{+}} - \mu_{\Sigma^{-}} + \mu_{\Xi^{0}} - \mu_{\Xi^{-}}}_\text{3.003 $\mu_N$ }$ [Sachs sum rule]
		\item $\displaystyle \underbrace{\mu_{\Sigma^{+}} + \mu_{\Sigma^{-}}}_\text{1.395 $\mu_N$ } = \underbrace{2 \mu_{\Sigma^{0}}}_\text{1.396 $\mu_N$ } = \underbrace{\frac{4}{3} (\mu_{p} + \mu_{n}) - \frac{2}{3} \mu_{\Lambda^{0}}}_\text{1.695 $\mu_N$}$
		\begin{enumerate}
			\item $s \to c$
			\newline $\displaystyle \underbrace{\mu_{\Sigma_{c}^{++}} + \mu_{\Sigma_{c}^{0}} }_\text{1.066 $\mu_N$ } = \underbrace{2 \mu_{\Sigma_{c}^{+}}}_\text{1.068 $\mu_N$ } = \underbrace{\frac{4}{3} (\mu_{p} + \mu_{n}) - \frac{2}{3} \mu_{\Lambda_{c}^{+}}}_\text{1.104 $\mu_N$}$
			\item $u \to s$, $s \to c$
			\newline$\displaystyle \underbrace{\mu_{\Omega_{c}^{0}} + \mu_{\Sigma_{c}^{0}}}_\text{-2.272 $\mu_N$ } = \underbrace{2 \mu_{\Xi_{c}^{\prime 0}}}_\text{-2.314 $\mu_N$ } = \underbrace{\frac{4}{3} (\mu_{\Xi^{-}} + \mu_{\Sigma^{-}}) - \frac{2}{3} \mu_{\Xi_{c}^{0}}}_\text{-2.592 $\mu_N$}$
			\item $d \to s$, $s \to c$
			\newline $\displaystyle \underbrace{\mu_{\Sigma_{c}^{++}} + \mu_{\Omega_{c}^{0}}}_\text{1.392 $\mu_N$ } = \underbrace{2 \mu_{\Xi_{c}^{\prime +}}}_\text{1.376 $\mu_N$ } = \underbrace{\frac{4}{3} (\mu_{\Sigma^{+}} + \mu_{\Xi^{0}}) - \frac{2}{3} \mu_{\Xi_{c}^{+}}}_\text{1.404 $\mu_N$}$
		\end{enumerate}
		\item $\displaystyle \underbrace{\mu_{\Xi^{*0}} - \mu_{\Xi^{*-}}}_\text{2.804 $\mu_N$ } = \underbrace{\mu_{\Sigma^{*0}} - \mu_{\Sigma^{*-}}}_\text{2.780 $\mu_N$} = \underbrace{\mu_{\Sigma^{*+}} - \mu_{\Sigma^{*0}}}_\text{2.777 $\mu_N$ } = \underbrace{\mu_{\Delta^{+}} - \mu_{\Delta^{0}}}_\text{2.752 $\mu_N$}$ 
		\begin{enumerate}
			\item $s \to c$
			\newline $\displaystyle \underbrace{\mu_{\Xi_{c}^{*+}} - \mu_{\Xi_{c}^{*0}} }_\text{2.673 $\mu_N$ } = \underbrace{\mu_{\Sigma_{c}^{*+}} - \mu_{\Sigma_{c}^{*0}}}_\text{2.656 $\mu_N$} = \underbrace{\mu_{\Sigma_{c}^{*++}} - \mu_{\Sigma_{c}^{*+}}}_\text{2.654 $\mu_N$ }  =\underbrace{\mu_{\Xi_{cc}^{*++}} - \mu_{\Xi_{cc}^{*+}} }_\text{2.546 $\mu_N$ }$ 
			\item $u \to s$, $s \to c$
			\newline $\displaystyle \underbrace{\mu_{\Omega_{cc}^{*+}} - \mu_{\Xi_{cc}^{*+}} }_\text{0.246 $\mu_N$ } = \underbrace{\mu_{\Xi_{c}^{*0}} - \mu_{\Sigma_{c}^{*0}}}_\text{0.199 $\mu_N$} = \underbrace{\mu_{\Omega_{c}^{*0}} - \mu_{\Xi_{c}^{*0}}}_\text{0.259 $\mu_N$ }  =\underbrace{\mu_{\Xi^{*-}} - \mu_{\Sigma^{*-}} }_\text{0.204 $\mu_N$ }$
			\item $d \to s$, $s \to c$
			\newline $\displaystyle \underbrace{\mu_{\Xi_{cc}^{*++}} - \mu_{\Omega_{cc}^{*+}} }_\text{2.300 $\mu_N$ } = \underbrace{\mu_{\Xi_{c}^{*+}} - \mu_{\Omega_{c}^{*0}}}_\text{2.415 $\mu_N$} = \underbrace{\mu_{\Sigma_{c}^{*++}} - \mu_{\Xi_{c}^{*+}}}_\text{2.437 $\mu_N$ }  =\underbrace{\mu_{\Sigma^{*+}} - \mu_{\Xi^{*0}} }_\text{2.548 $\mu_N$ }$
		\end{enumerate}
		\item $\displaystyle \underbrace{\mu_{\Sigma^{+}} - \mu_{\Sigma^{-}}}_\text{3.755 $\mu_N$ } = \underbrace  { -4 (\mu_{\Xi^{0}} - \mu_{\Xi^{-}})}_\text{3.009 $\mu_N$}$
		\begin{enumerate}
			\item $s \to c$
			\newline $\displaystyle \underbrace{\mu_{\Sigma_{c}^{++}} - \mu_{\Sigma_{c}^{0}}}_\text{3.665 $\mu_N$ } = \underbrace  { 2 (\mu_{\Xi_{c}^{\prime +}} - \mu_{\Xi_{c}^{\prime 0}})}_\text{3.690 $\mu_N$} $\ \ \ \ \  (here, only one $s$ quark transforms to $c$ quark)
			\newline $\displaystyle \underbrace{\mu_{\Sigma_{c}^{++}} - \mu_{\Sigma_{c}^{0}}}_\text{3.665 $\mu_N$ } = \underbrace  { -4 (\mu_{\Xi_{cc}^{++}} - \mu_{\Xi_{cc}^{+}})}_\text{3.613 $\mu_N$} $\ \ \ \ \  (here, all $s$ quarks transform to $c$ quarks)
			\item $u \to s$, $s \to c$
			\newline $\displaystyle \underbrace{\mu_{\Omega_{c}^{0}} - \mu_{\Sigma_{c}^{0}}}_\text{0.326 $\mu_N$ } = \underbrace  { -4 (\mu_{\Omega_{cc}^{+}} - \mu_{\Xi_{cc}^{+}})}_\text{0.376 $\mu_N$}$
			\item $d \to s$, $s \to c$
			\newline $\displaystyle \underbrace{\mu_{\Sigma_{c}^{++}} - \mu_{\Omega_{c}^{0}}}_\text{3.339 $\mu_N$ } = \underbrace  { -4 (\mu_{\Xi_{cc}^{++}} - \mu_{\Omega_{cc}^{+}})}_\text{3.237 $\mu_N$}$
		\end{enumerate}
		\item $\displaystyle \underbrace{\mu_{\Xi^{*0}} +  \mu_{\Xi^{*-}}}_\text{-1.982 $\mu_N$} = \underbrace{\mu_{\Sigma^{*0}} + \mu_{\Omega^{-}}}_\text{-1.948 $\mu_N$ }$
		\begin{enumerate}
			\item $s \to c$
			\newline $\displaystyle \underbrace{\mu_{\Xi_{c}^{*+}} + \mu_{\Xi_{c}^{* 0}}}_\text{0.434 $\mu_N$} = \underbrace{\mu_{\Sigma_{c}^{*+}} +\mu_{\Omega_{c}^{*0}}}_\text{0.476 $\mu_N$ }$\ \ \ \ \  (here, only one $s$ quark transforms to $c$ quark)
			\newline $\displaystyle \underbrace{\mu_{\Xi_{cc}^{*++}} + \mu_{\Xi_{cc}^{*+}}}_\text{2.541 $\mu_N$} = \underbrace{\mu_{\Sigma_{c}^{*+}} +\mu_{\Omega_{ccc}^{*++}}}_\text{2.516 $\mu_N$ }$\ \ \ \ \  (here, all $s$ quarks transform to $c$ quarks)
			\item $u \to s$, $s \to c$ 
			\newline $\displaystyle \underbrace{\mu_{\Omega_{cc}^{*+}} + \mu_{\Xi_{cc}^{*+}}}_\text{0.241 $\mu_N$} = \underbrace{\mu_{\Xi_{c}^{*0}} +\mu_{\Omega_{ccc}^{*++}}}_\text{0.060 $\mu_N$ }$
			\item $d \to s$, $s \to c$ 
			\newline $\displaystyle \underbrace{\mu_{\Xi_{cc}^{*++}} + \mu_{\Omega_{cc}^{*+}}}_\text{2.787 $\mu_N$} = \underbrace{\mu_{\Xi_{c}^{*+}} +\mu_{\Omega_{ccc}^{*++}}}_\text{2.733 $\mu_N$ }$
		\end{enumerate}
		\item $\displaystyle \underbrace{4 (\mu_{p} - \mu_{n})}_\text{18.64 $\mu_N$} =\underbrace{5 (\mu_{\Sigma^{+}} - \mu_{\Sigma^{-}})}_\text{18.78 $\mu_N$ } $ 
		\begin{enumerate}
			\item $s \to c$
			\newline $\displaystyle \underbrace{4 (\mu_{p} - \mu_{n})}_\text{18.64 $\mu_N$} =\underbrace{5 (\mu_{\Sigma_{c}^{++}} - \mu_{\Sigma_{c}^{0}})}_\text{18.32 $\mu_N$ }$ 
			\item $u \to s$, $s \to c$
			\newline $\displaystyle \underbrace{4 (\mu_{\Xi^{-}} - \mu_{\Sigma^{-}})}_\text{2.408 $\mu_N$} =\underbrace{5 (\mu_{\Omega_{c}^{0}} - \mu_{\Sigma_{c}^{0}})}_\text{1.630 $\mu_N$ }$
			\item $d \to s$, $s \to c$
			\newline $\displaystyle \underbrace{4 (\mu_{\Sigma^{+}} - \mu_{\Xi^{0}})}_\text{15.62 $\mu_N$} =\underbrace{5 (\mu_{\Sigma_{c}^{++}} - \mu_{\Omega_{c}^{0}})}_\text{16.69 $\mu_N$ }$
		\end{enumerate}
		\item $\displaystyle \underbrace{\mu_{\Delta^{+}} - \mu_{\Delta^{-}}}_\text{5.507 $\mu_N$} =\underbrace{\mu_{\Sigma^{*+}} - \mu_{\Sigma^{*-}}}_\text{5.557 $\mu_N$ } $ 
		\begin{enumerate}
			\item $s \to c$
			\newline $\displaystyle \underbrace{\mu_{\Delta^{+}} - \mu_{\Delta^{-}}}_\text{5.507 $\mu_N$} =\underbrace{\mu_{\Sigma_{c}^{*++}} - \mu_{\Sigma_{c}^{*0}}}_\text{5.310 $\mu_N$ }$
			\item $u \to s$, $s \to c$
			\newline $\displaystyle \underbrace{\mu_{\Xi^{*-}} - \mu_{\Delta^{-}}}_\text{0.354 $\mu_N$} =\underbrace{\mu_{\Omega_{c}^{*0}} - \mu_{\Sigma_{c}^{*0}}}_\text{0.458 $\mu_N$ }$
			\item $d \to s$, $s \to c$
			\newline $\displaystyle \underbrace{\mu_{\Sigma^{*+}} - \mu_{\Omega^{-}}}_\text{5.090 $\mu_N$} =\underbrace{\mu_{\Sigma_{c}^{*++}} - \mu_{\Omega_{c}^{*0}}}_\text{4.852 $\mu_N$ }$
		\end{enumerate}
		\item $\displaystyle \underbrace{\mu_{\Delta^{+}} - \mu_{\Sigma^{*+}} }_\text{-0.199 $\mu_N$ } = \underbrace{\mu_{\Xi^{*-}} - \mu_{\Omega^{-}} }_\text{-0.262 $\mu_N$ } $  
		\begin{enumerate}
			\item $s \to c$
			\newline $\displaystyle \underbrace{\mu_{\Delta^{+}} - \mu_{\Sigma_{c}^{*++}} }_\text{-1.231 $\mu_N$ } = \underbrace{\mu_{\Xi_{cc}^{*+}} - \mu_{\Omega_{ccc}^{*++}} }_\text{-1.182 $\mu_N$ } $
			\item $u \to s$, $s \to c$
			\newline $\displaystyle \underbrace{\mu_{\Xi^{*-}} - \mu_{\Omega_{c}^{*0}} }_\text{-1.532 $\mu_N$ } = \underbrace{\mu_{\Xi_{cc}^{*+}} - \mu_{\Omega_{ccc}^{*++}} }_\text{-1.182 $\mu_N$ } $
			\item $d \to s$, $s \to c$
			\newline $\displaystyle \underbrace{\mu_{\Sigma^{*+}} - \mu_{\Sigma_{c}^{*++}}}_\text{-1.032 $\mu_N$ } = \underbrace{\mu_{\Omega_{cc}^{*+}} - \mu_{\Omega_{ccc}^{*++}} }_\text{-0.936 $\mu_N$ } $
		\end{enumerate}
		\item $\displaystyle \underbrace{\mu_{\Delta^{++}} - \mu_{\Delta^{-}}}_\text{8.258 $\mu_N$} =\underbrace{3 (\mu_{\Delta^{+}} - \mu_{\Delta^{0}})}_\text{8.257 $\mu_N$ }$ 
	\end{enumerate}
	
	%%%%%%%%%%%%%%%%%%%%%%%%%%%%%%%%%%%%%%%%%%%%%%%%%%%%%%%%%%%%%%%%%%%%%%%%%%%%%%%%% 
	\newpage
	\begin{table}[ht]
		\centering
		\captionof{table}{Masses of light baryons (in MeV).} 
		\label{mass_light}
		\begin{tabular}{|c|c|c|c|}	\hline  
			\textbf{$SU(3)$} &\multirow{2}{*}{\textbf{Baryons}} & \textbf{Our} & \textbf{PDG}\symbolfootnote[3]{Values in the parentheses represent uncertainties.} \\ 
			\textbf{multiplet} & & \textbf{work} & \bf\cite{Workman:2022ynf} \\
			\hline	\hline
			\multicolumn{4}{|l|}{$J^P=\frac{1}{2}^+$}\\ \hline
			\multirow{7}{*}{Octet}
			&$p$           & $936.94$ \symbolfootnote[4]{Used as input in the calculation of constituent quark masses ($m_{i}$).}$^{,}$\symbolfootnote[5]{Used as input in the calculation of hyperfine interaction terms ($b_{ij}$).}  & $938.27$    \\	
			\multirow{7}{*}{(C = 0)}&$n$           & $938.24$ \footnotemark[4]$^{,}$\footnotemark[5] & $939.56$  \\
			&$\Lambda^{0}$ & $1115.68$ \footnotemark[4]& $1115.683(6)$  \\	
			&$\Sigma^{+}$  & $1168.04$ & $1189.37(7)$  \\	
			&$\Sigma^{0}$  & $1172.25$ & $1192.642(24)$ \\		
			&$\Sigma^{-}$  & $1176.19$ & $1197.449(30)$ \\
			&$\Xi^{0}$     & $1314.86$ \footnotemark[5] & $1314.86(20)$  \\
			&$\Xi^{-}$     & $1321.71$ \footnotemark[5] & $1321.71(7)$  \\\hline \hline
			\multicolumn{4}{|l|}{$J^P=\frac{3}{2}^+$}\\ \hline
			\multirow{9}{*}{Decuplet}
			&$\Delta^{++}$ & $1232.00$ \footnotemark[4]$^{,}$\footnotemark[5] & $1230.55(20)$ \\
			\multirow{9}{*}{(C = 0)}&$\Delta^{+}$  & $1233.57$ \footnotemark[4]$^{,}$\footnotemark[5] & $1234.9(1.4)$ \\
			&$\Delta^{0}$  & $1234.86$ \footnotemark[4]$^{,}$\footnotemark[5] & $1231.3(6)$ \\
			&$\Delta^{-}$  & $1235.89$ & -  \\
			&$\Sigma^{*+}$ & $1382.74$ & $1382.83(34)$ \\	
			&$\Sigma^{*0}$ & $1384.03$ & $1383.7(1.0)$  \\		
			&$\Sigma^{*-}$ & $1385.04$ & $1387.2(5)$  \\
			&$\Xi^{*0}$    & $1529.55$ & $1531.80(32)$ \\
			&$\Xi^{*-}$    & $1530.56$ & $1535.0(6)$  \\
			&$\Omega^{-}$  & $1672.45$ \footnotemark[5] & $1672.45(29)$  \\\hline
		\end{tabular}
	\end{table}
	%%%%%%%%%%%%%%%%%%%%%%%%%%%%%%%%%%%%%%%%%%%%%%%%%%%%%%%%%%%%%%%%%%%%%%%%%%  
	\begin{table}[ht]
		\centering
		\captionof{table}{Masses of charm baryons (in MeV).} 
		\label{mass_charm}
		\begin{tabular}{|c|c|c|c|c|c|c|c|}	\hline  
			\textbf{$SU(3)$} &\multirow{2}{*}{\textbf{Baryons}} &\textbf{Our}
			& \textbf{HB$\chi$PT} & \textbf{hCQM} & \textbf{QCDSR}  & \textbf{CI} & \textbf{PDG}\\ 
			\textbf{multiplet} & &\textbf{work}
			&\bf\cite{Jiang:2014ena} &\bf\cite{Shah:2016mig, Shah:2016vmd, Shah:2017liu, Shah:2017jkr} & \bf\cite{Zhang:2009iya} & \bf\cite{Gutierrez-Guerrero:2019uwa} & \bf\cite{Workman:2022ynf}\\ 
			\hline	\hline
			\multicolumn{8}{|l|}{$J^P=\frac{1}{2}^+$}\\ \hline
			\multirow{2}{*}{Anti-triplet}
			&$\Lambda_c^{+}$      & $2220.59$  & $2286.46$ \footnotemark[6] & $2286$ \footnotemark[6] & $2310$ & -      & $2286.46(14)$  \\
			\multirow{2}{*}{(C = 1)}&$\Xi_c^{+}$          & $2438.04$  & $2467.80$ \footnotemark[6] & $2467$ \footnotemark[6] & -      & $2700$ & $2467.71(23)$ \\	
			&$\Xi_c^{0}$          & $2443.91$  & $2470.88$ \footnotemark[6] & $2470$ \footnotemark[6] & $2480$ & -      & $2470.44(28)$ \\\hline 
			
			\multirow{5}{*}{Sextet}
			&$\Sigma_c^{++}$      & $2374.01$  & $2454.02$ \footnotemark[6] & $2454$ \footnotemark[6] & -      & $2580$ & $2453.97(14)$ \\
			\multirow{5}{*}{(C = 1)}&$\Sigma_c^{+}$       & $2375.90$  & $2452.90$ \footnotemark[6] & $2452$ \footnotemark[6] & -      & -      & $2452.65^{+0.22}_{-0.16}$ \\
			&$\Sigma_c^{0}$       & $2377.52$  & $2453.76$ \footnotemark[6] & $2453$ \footnotemark[6] & $2400$ & -      & $2453.75(14)$  \\	
			&$\Xi_c^{\prime+}$    & $2536.57$ \footnotemark[5]  & $2572.66$ & -      & -      & -      & $2578.2(5)$\\
			&$\Xi_c^{\prime0}$    & $2538.18$ \footnotemark[5]  & $2570.40$ & -      & $2500$ & -      & $2578.7(5)$  \\
			&$\Omega_c^{0}$       & $2695.20$ \symbolfootnote[4]{Used as input in the calculation of constituent quark masses ($m_{i}$).}$^{,}$\symbolfootnote[5]{Used as input in the calculation of hyperfine interaction terms ($b_{ij}$).}  & $2695.20$ \footnotemark[6] & $2695$ \symbolfootnote[6]{Used as input in respective models.} & $2620$ & $2820$ & $2695.2(1.7)$ \\ \hline
			
			\multirow{2}{*}{Triplet}
			&$\Xi_{cc}^{++}$      & $3621.60$ \footnotemark[5] & $3665$ \bf\cite{Sun:2014aya} & $3511$ & -      & $3640$ & $3621.6(4)$ \\
			\multirow{2}{*}{(C = 2)}&$\Xi_{cc}^{+}$       & $3623.81$  & - & $3520$ & -      & -      & $3623.0(1.4)$ \bf\cite{LHCb:2021eaf} \\
			&$\Omega_{cc}^{+}$    & $3795.97$  & - & $3650$ & $4250$ & $3760$ & $3738$ \textbf{(LQCD \cite{Brown:2014ena})} \\ \hline \hline
			\multicolumn{8}{|l|}{$J^P=\frac{3}{2}^+$}\\ \hline
			\multirow{5}{*}{Sextet}
			&$\Sigma_c^{*++}$     & $2437.11$  & $2518.40$ \footnotemark[6] & $2530$ & & $2810$ & $2518.41^{+0.22}_{-0.18}$ \\
			\multirow{5}{*}{(C = 1)}&$\Sigma_c^{*+}$      & $2439.56$  & $2517.50$ \footnotemark[6] & $2501$ & -      & -      & $2517.4^{+0.7}_{-0.5}$ \\
			&$\Sigma_c^{*0}$      & $2441.74$  & $2518.0$ \footnotemark[6]  & $2529$ & $2560$ & -      & $2518.48(20)$ \\	
			&$\Xi_c^{*+}$         & $2603.47$ \footnotemark[5] & $2636.83$ & $2619$ & -      & -      & $2645.10(30)$ \\	
			&$\Xi_c^{*0}$         & $2605.64$ \footnotemark[5] & $2633.71$ & $2610$ & $2640$ & -      & $2646.16(25)$ \\		
			&$\Omega_c^{*0}$      & $2765.90$ \footnotemark[4] & $2765.90$ \footnotemark[6] & $2745$ & $2740$ & $3007$ & $2765.9(2.0)$ \\ \hline

			\multirow{2}{*}{Triplet}
			&$\Xi_{cc}^{*++}$     & $3684.70$  & $3726$ \bf\cite{Sun:2014aya} & $3687$ & -      & $3895$ & $3692$ \textbf{(LQCD \cite{Brown:2014ena})}  \\
			\multirow{2}{*}{(C = 2)}&$\Xi_{cc}^{*+}$      & $3688.03$  & - & $3695$ & -      & -      & - \\
			&$\Omega_{cc}^{*+}$   & $3866.67$  & - & $3810$ & $3810$ & $4043$ & $3822$ \textbf{(LQCD \cite{Brown:2014ena})} \\ \hline
			
			\multirow{1}{*}{Singlet (C = 3)}
			&$\Omega_{ccc}^{*++}$ & $4974.76$  & - & $4806$ & $4670$ & $4930$ & $4796$ \textbf{(LQCD \cite{Brown:2014ena})} \\ \hline
		\end{tabular}
	\end{table}
	%%%%%%%%%%%%%%%%%%%%%%%%%%%%%%%%%%%%%%%%%%%%%%%%%%%%%%%%%%%%%%%%%%%%%%%%%%%%%%%%
	\newpage
	\begin{table}
		\centering
		\captionof{table}{Isospin mass splittings of light baryons (in MeV).} 
		\label{isospin_strange}
		\begin{tabular}{|c|c|c|}	\hline  
			\textbf{Splitting} & \textbf{Our work} & \textbf{PDG \cite{Workman:2022ynf}}\\	\hline	\hline
			$M_{n} -  M_{p}$                     & $1.294$ \symbolfootnote[7]{Inputs.} & $1.293$           \\	
			$M_{\Sigma^{-}} -  M_{\Sigma^{+}}$   & $8.14$  & $8.08(8)$   \\
			$M_{\Sigma^{-}} -  M_{\Sigma^{0}}$   & $3.94$  & $4.807(35)$ \\	
			$M_{\Sigma^{0}} -  M_{\Sigma^{+}}$   & $4.2$   & -                 \\	
			$M_{\Xi^{-}} - M_{\Xi^{0}}$          & $6.85$  & $6.85(21)$   \\	\hline	
			$M_{\Delta^{0}} - M_{\Delta^{++}}$   & $2.86$ \footnotemark[7] & $2.86(30)$   \\
			$M_{\Sigma^{*-}} -  M_{\Sigma^{*+}}$ & $2.3$   & -                 \\
			$M_{\Sigma^{*-}} -  M_{\Sigma^{*0}}$ & $1.01$  & -                 \\		
			$M_{\Sigma^{*0}} -  M_{\Sigma^{*+}}$ & $1.29$  & -                 \\
			$M_{\Xi^{*-}} - M_{\Xi^{*0}}$        & $1.01$  & $3.2(6)$\symbolfootnote[8]{The numerical value of mass splitting ranges from $2~ \text{to}~7$ MeV with larger uncertainties \cite{Workman:2022ynf}.}  \\	\hline
		\end{tabular}
	\end{table}
	%%%%%%%%%%%%%%%%%%%%%%%%%%%%%%%%%%%%%%%%%%%%%%%%%%%%%%%%%%%%%%%%%%%%%%%%%%%%%%%%%%%%
	\begin{table}
		\centering
		\captionof{table}{Isospin mass splittings of charm baryons (in MeV).} 
		\label{isospin_charm}
		\begin{tabular}{|c|c|c|}
			\hline 
			\textbf{Splitting} & \textbf{Our work} & \textbf{pion mean-field}\symbolfootnote[9]{We have taken isospin mass splitting as, $\Delta M^{iso}_{sb} + \Delta M_{hf}$, following the notations from \cite{Yang:2020klp}.} \bf\cite{Yang:2020klp} \\	\hline	\hline
			$M_{\Sigma_{c}^{0}} -  M_{\Sigma_{c}^{++}}$     & $3.51$ & $3.86$  \\	
			$M_{\Sigma_{c}^{+}} -  M_{\Sigma_{c}^{++}}$     & $1.89$ & $1.93$  \\
			$M_{\Sigma_{c}^{0}} -  M_{\Sigma_{c}^{+}}$      & $1.62$ & $1.93$  \\	
			$M_{\Xi_{c}^{0}} - M_{\Xi_{c}^{+}}$             & $5.87$ & $4.71$  \\	
			$M_{\Xi_{c}^{\prime0}} - M_{\Xi_{c}^{\prime+}}$ & $1.61$ & $1.93$  \\ \hline	
			$M_{\Sigma_{c}^{*0}} -  M_{\Sigma_{c}^{*++}}$   & $4.63$ & $3.86$  \\	
			$M_{\Sigma_{c}^{*+}} -  M_{\Sigma_{c}^{*++}}$   & $2.45$ & $1.93$  \\	
			$M_{\Sigma_{c}^{*0}} -  M_{\Sigma_{c}^{*+}}$    & $2.18$ & $1.93$  \\	
			$M_{\Xi_{c}^{*0}} - M_{\Xi_{c}^{*+}}$           & $2.17$ & $1.93$  \\	\hline 
		\end{tabular}
	\end{table}
	%%%%%%%%%%%%%%%%%%%%%%%%%%%%%%%%%%%%%%%%%%%%%%%%%%%%%%%%%%%%%%%%%%%
	\begin{table}
		\centering
		\captionof{table}{Isospin mass splittings of doubly charm baryons (in MeV).} 
		\label{isospin_doubly_charm}
		\begin{tabular}{|c|c|c|}
			\hline 
			\textbf{Splitting} &$M_{\Xi_{cc}^{+}} - M_{\Xi_{cc}^{++}}$ & $M_{\Xi_{cc}^{*+}} - M_{\Xi_{cc}^{*++}}$         \\	\hline  \hline
			Our work                                               & $2.21$                    & $3.33$              \\
			HQS \cite{Hwang:2008dj}                                & $-2.3(1.7)$           &          -          \\
			Z. Shah \cite{Shah:2017liu}                           & $9$ &          -          \\
			Ke-Wei Wei \textit{et al.} \cite{Wei:2015gsa}          & $-0.4(3)$             & $-1.3^{+1.1}_{-1.2}$ \\
			LQCD \cite{Borsanyi:2014jba}                           & $-2.16(11)(17)$ &          -          \\
			Karliner \textit{et al.} \cite{Karliner:2017gml}    & $-2.17(11)$          &          -          \\
			Brodsky \textit{et al.} \cite{Brodsky:2011zs}    & $-1.5(2.7)$            &          -          \\ \hline
			
		\end{tabular}
	\end{table}
	%%%%%%%%%%%%%%%%%%%%%%%%%%%%%%%%%%%%%%%%%%%%%%%%%%%%%%%%%%%%%%%%%%
	\begin{table}
		\centering
		\captionof{table} {Magnetic moments of $J^P=\frac{1}{2}^+$ light baryons (in unit of nuclear magneton, $\mu_N$).}
		\label{octet_light_mm}
		\begin{tabular}{|c|c|c|c|}	\hline
			\textbf{Baryons} & \textbf{EMS}& \textbf{SQCS } &\textbf{PDG \cite{Workman:2022ynf}}\\ \hline    \hline
			\multicolumn{4}{|l|}{Octet C = 0}\\\hline
			$p$            & $2.875$  & $2.840$  & $2.793$                \\	
			$n$            & $-2.030$ & $-1.820$ & $-1.913$               \\	
			$\Lambda^{0}$  & $-0.579$ & $-0.502$ & $-0.613(4)$     \\
			$\Sigma^{+}$   & $2.607$  & $2.575$  & $2.458(10)$      \\		
			$\Sigma^{0}$   & $0.825$  & $0.698$  & $0.649(13)$ \symbolfootnote[10]{Calculated according to the definition $\mu_{\Sigma^{0}} = \frac{1}{2}(\mu_{\Sigma^{+}}^{Expt.} + \mu_{\Sigma^{-}}^{Expt.})$ \cite{Barik:1986fp, Tsushima:2020gun}.}\\	
			$\Sigma^{-}$   & $-0.960$ & $-1.180$ & $-1.160(25)$     \\	
			$\Xi^{0}$      & $-1.534$ & $-1.330$ & $-1.250(14)$     \\	
			$\Xi^{-}$      & $-0.457$ & $-0.578$ & $-0.6507(25)$   \\ \hline
		\end{tabular}
	\end{table}		
	%%%%%%%%%%%%%%%%%%%%%%%%%%%%%%%%%%%%%%%%%%%%%%%%%%%%%%%%%%%%%%%%%%%%%%%%%%%%%%%%%%%%
	\begin{table}
		\centering
		\captionof{table} {Magnetic moments of $J^P=\frac{3}{2}^+$ light baryons (in $\mu_N$).}
		\label{decuplet_light_mm}
		\begin{tabular}{|c|c|c|c|c|c|c|}	\hline 
			\textbf{Baryons} & \textbf{EMS}& \textbf{SQCS} & \textbf{$\chi$QM \cite{Linde:1997ni}}  & \textbf{$\chi$CQM \cite{Sharma:2010vv}}  & \textbf{HCM \cite{Ghalenovi:2017fvw}}  &\textbf{PDG \cite{Workman:2022ynf}} \\\hline	\hline
			\multicolumn{7}{|l|}{Decuplet C = 0}\\\hline
			$\Delta^{++}$ & $4.569$  & $5.511$  & $5.300$  & $4.510$  & $4.55$  & $6.14(51)$ \\
			$\Delta^{+}$  & $2.291$  & $2.760$  & $2.580$  & $2.000$  & $2.27$  & $2.7(1.5)$ \\	
			$\Delta^{0}$  & $0.009$  & $0.008$  & $-0.130$ & $-0.510$ & $0.00$  & -          \\		
			$\Delta^{-}$  & $-2.278$ & $-2.747$ & $-2.850$ & $-3.020$ & $-2.27$ & -          \\
			$\Sigma^{*+}$ & $2.557$  & $2.959$  & $2.880$  & $2.690$  & $2.72$  & -          \\	
			$\Sigma^{*0}$ & $0.237$  & $0.182$  & $0.170$  & $0.020$  & $0.27$  & -          \\ 	
			$\Sigma^{*-}$ & $-2.088$ & $-2.597$ & $-2.550$ & $-2.640$ & $-2.17$ & -          \\ 	
			$\Xi^{*0}$    & $0.474$  & $0.411$  & $0.470$  & $0.540$  & $0.57$  & -          \\	
			$\Xi^{*-}$    & $-1.890$ & $-2.393$ & $-2.250$ & $-1.870$ & $-1.96$ & -          \\
			$\Omega^{-}$  & $-1.683$ & $-2.131$ & $-1.950$ & $-1.710$ & $-1.70$ & $-2.02(5)$ \\ \hline 
		\end{tabular}
	\end{table}
	
	%%%%%%%%%%%%%%%%%%%%%%%%%%%%%%%%%%%%%%%%%%%%%%%%%%%%%%%%%%%%%%%%%%%%%%%%%%%%
	%%%%%%%%%%%%%%%%%%%%%%%%%%%%%%%%%%%%%%
	\begin{sidewaystable}
		\caption {Magnetic moments of $J^P=\frac{1}{2}^+$ charm baryons (in $\mu_N$).}
		\label{octet_charm_mm}
		\begin{tabular}{|c|c|c|c|c|c|c|c|c|c|c|c|}	\hline  
			\multirow{2}{*}{\textbf{Baryons}} & \multirow{2}{*}{\textbf{EMS}} & \multicolumn{2}{c|}{\textbf{SQCS}} & \textbf{BM} & \textbf{$\chi$CQM} & \textbf{NRQM} & \textbf{RTQM} & \textbf{HB$\chi$PT} & \textbf{B$\chi$PT} & \textbf{LCQSR} & \textbf{LQCD}\\
			\cline{3-4} 
			& & \textbf{$z =0.021$}& \textbf{$z =0.155$} & \bf\cite{Simonis:2018rld}& \bf\cite{Sharma:2010vv}& \bf\cite{Bernotas:2012nz}& \bf\cite{Faessler:2006ft} & \bf\cite{Wang:2018gpl}& \bf\cite{Shi:2018rhk, Liu:2018euh} & \bf\cite{Aliev:2015axa, Aliev:2001ig, Aliev:2008ay, Ozdem:2018uue}& \bf\cite{Can:2021ehb}\\   \hline	\hline
			\multicolumn{12}{|l|}{Sextet C = 1}\\\hline
			$\Sigma_c^{++}$   & $2.095$  & $ 2.365$ &  $2.622$  & $2.280$  & $2.200$  & $2.350$  & $1.760$  & $1.50$  & $2.00$  & $2.40$    & $2.220(505)$   \\	
			$\Sigma_c^{+}$    & $0.432$  & $0.534$  &  $0.818$  & $0.487$  & $0.300$  & $0.490$  & $0.360$  & $0.30$  & $0.46$  & $0.50$    & -              \\	
			$\Sigma_c^{0}$    & $-1.234$ & $-1.299$ &  $-0.987$ &$-1.310$  & $-1.600$ & $-1.370$ & $-1.040$ & $-0.91$ & $-1.08$ & $-1.50$   & $-1.073 (269)$ \\
			$\Xi_c^{\prime+}$ & $0.623$  & $0.688$  &  $0.927$  & $0.633$  & $0.760$  & $0.890$  & $0.470$  & $-0.31$ & $0.62$  & $0.80$    & -              \\
			$\Xi_c^{\prime0}$ & $-1.074$ & $-1.157$ &  $-0.889$ & $-1.120$ & $-1.320$ & $-1.180$ & $-0.950$ & $-0.80$ & $-0.91$ & $-1.20$   & -              \\   	
			$\Omega_c^{0}$    & $-0.905$ & $-0.973$ &  $-0.751$ & $-0.950$ & $-0.900$ & $-0.940$ & $-0.850$ & $-0.69$ & $-0.74$ & $-0.90$   & $-0.639(88)$   \\ \hline
			\multicolumn{12}{|l|}{Anti-triplet C = 1}\\\hline
			$\Lambda_c^{+}$   & $0.380$  & $ 0.384$ & $0.410$   & $0.335$  & $0.392$  & $0.390$  & $0.420$  & $0.24$  & $0.24$  & $0.40(5)$  & -              \\
			$\Xi_c^{+}$       & $0.380$  & $0.384$  &  $0.410$  & $0.334$  & $0.400$  & $0.200$  & $0.410$  & $0.29$  & $0.24$  & $0.50(5)$  & -              \\
			$\Xi_c^{0}$       & $0.380$  & $0.372$  &  $0.321$  & $0.334$  & $0.280$  & $0.410$  & $0.390$  & $0.19$  & $0.19$  & $0.35(5)$  & -              \\\hline	
			\multicolumn{12}{|l|}{Triplet C = 2}\\\hline
			$\Xi_{cc}^{++}$   & $-0.106$ & $-0.110$ &  $-0.137$ & $-0.110$ & $ 0.006$ & $-0.100$ & $0.130$  & -       & -       & $-0.23(5)$ & -              \\	
			$\Xi_{cc}^{+}$    & $0.813$  & $ 0.793$ &  $0.664$  & $0.719$  & $0.840$  & $0.830$  & $0.720$  & -       & $0.39$  & $0.43(9)$  & $0.425(29)$    \\
			$\Omega_{cc}^{+}$ & $0.710$  & $0.699$  &  $0.625$  & $0.645$  & $0.697$  & $0.720$  & $0.670$  & -       & $0.39$  & $0.39(9)$  & $0.413(24)$    \\ \hline 
		\end{tabular}
	\end{sidewaystable}
	%%%%%%%%%%%%%%%%%%%%%%%%%%%%%%%%%%%%%%%%%%%%%%%%%%%%%
	
	\begin{table}
		\caption{Magnetic moments of $J^P=\frac{3}{2}^+$ charm baryons (in $\mu_N$).}
		\label{decuplet_charm_mm}
		\begin{tabular}{|c|c|c|c|c|c|c|c|c|c|c|}	\hline 
			\multirow{2}{*}{\textbf{Baryons}} & \multirow{2}{*}{\textbf{EMS}} & \multicolumn{2}{c|}{\textbf{SQCS}} & \textbf{BM} & \textbf{$\chi$CQM} & \textbf{NRQM} & \textbf{HCM} & \textbf{HB$\chi$PT} & \textbf{B$\chi$PT} & \textbf{LCQSR}\\
			\cline{3-4} 
			& &\textbf{$z =0.021$}& \textbf{$z =0.155$}& \bf\cite{Simonis:2018rld} &\bf\cite{Sharma:2010vv} &\bf\cite{Bernotas:2012nz}&\bf\cite{Patel:2007gx, Patel:2008mv} & \bf\cite{Meng:2018gan} & \bf\cite{Shi:2021kmm} &\bf\cite{Aliev:2008sk, Ozdem:2019zis}\\\hline	\hline
			\multicolumn{11}{|l|}{Sextet C = 1}\\\hline
			$\Sigma_c^{*++}$      & $3.578$  & $3.991$  & $4.507$  & $3.980$  & $3.920$  & $4.110$  & $3.842$  & $2.410$  & -       & $4.81(1.22)$ \\ 	
			$\Sigma_c^{*+}$       & $1.185$  & $1.337$  & $1.782$  & $1.250$  & $0.970$  & $1.320$  & $1.252$  & $0.670$  & -       & $2.00(46)$   \\	
			$\Sigma_c^{*0}$       & $-1.214$ & $-1.318$ & $-0.944$ & $-1.490$ & $-1.990$ & $-1.470$ & $-0.848$ & $-1.070$ & -       & $-0.81(20)$  \\
			$\Xi_c^{*+}$          & $1.453$  & $1.554$  & $1.935$  & $1.470$  & $1.590$  & $1.640$  & $1.513$  & $0.810$  & -       & $1.68(42)$   \\
			$\Xi_c^{*0}$          & $-0.987$ & $-1.119$ & $-0.808$ & $-1.200$ & $-1.430$ & $-1.150$ & $-0.688$ & $-0.900$ & -       & $-0.68(18)$  \\
			$\Omega_c^{*0}$       & $-0.750$ & $-0.861$ & $-0.613$ & $-0.936$ & $-0.860$ & $-0.830$ & $-0.865$ & $-0.700$ & -       & $-0.62(18)$  \\	\hline
			\multicolumn{11}{|l|}{Triplet C = 2}\\\hline
			$\Xi_{cc}^{*++}$      & $2.441$  & $2.543$  & $3.189$  & $2.350$  & $2.660$  & $2.640$  &  $2.749$ & -        & $2.890$  & $2.94(95)$   \\
			$\Xi_{cc}^{*+}$       & $-0.081$ & $-0.003$ & $0.494$  & $-0.178$ & $-0.470$ & $-0.150$ & $-0.168$ & -        & $-0.250$ & $-0.67(11)$  \\	
			$\Omega_{cc}^{*+}$    & $0.188$  & $0.243$  & $0.598$  & $0.048$  & $0.140$  & $0.170$  & $0.121$  & -        & $0.001$ & $-0.52(7)$   \\	\hline
			\multicolumn{11}{|l|}{Singlet C = 3}\\\hline
			$\Omega_{ccc}^{*++}$  & $1.132$  & $1.179$  & $1.483$  & $0.989$  & $0.155$  & $1.170$  & $1.189$  &  -       & -       &-         \\	\hline 
		\end{tabular}
	\end{table}
	%%%%%%%%%%%%%%%%%%%%%%%%%%%%%%%%%%%%%%%%%%%%%%%%%%%%%%%%%%%%%%%%%%%%%%%%%%%%%%%%%%%%
	\begin{table}
		\centering
		\captionof{table}{Transition moments of light baryons (in $\mu_N$).} 
		\label{transition_light}
		\begin{tabular}{|c|c|c|c|c|c|c|c|c|}	\hline  
			\multirow{2}{*}{\textbf{Transitions}\symbolfootnote[11]{Experimental value for $\Sigma^{0}\rightarrow \Lambda^{0}$, $\Delta^{+}\rightarrow p$, $\Sigma^{*0}\rightarrow \Lambda^{0}$, and $\Sigma^{*+}\rightarrow \Sigma^{+}$  are $-1.61(8)~\mu_N$ \cite{Workman:2022ynf}, $3.51(9)~\mu_N$ \cite{Flores-Mendieta:2021yzz}, $2.75(25)~\mu_N$ \cite{CLAS:2011vzg}, and $3.17(36)~\mu_N$ \cite{CLAS:2011iuw}, respectively.}} & \multirow{2}{*}{\textbf{EMS}} & \multirow{2}{*}{\textbf{SQCS}} & \textbf{$\chi$CQM} & \textbf{HCM} & \textbf{$\chi$QM} & \textbf{CSQM} & \textbf{$\chi$PT} & \textbf{LCQSR} \\
			& & & \bf\cite{Sharma:2010vv} & \bf\cite{Ghalenovi:2017fvw} & \bf\cite{Yu:2006sc} & \bf\cite{Ramalho:2020tnn} & \bf\cite{Li:2017vmq} & \bf\cite{Aliev:2004ju} \\\hline \hline
			\multicolumn{9}{|l|}{(C = 0) Octet $\rightarrow$ Octet}\\\hline
			$\Sigma^{0}\rightarrow \Lambda^{0}$     & $-1.706$ & $-1.530$ & $1.600$  & -    & $1.682$  & -       & -       & - \\\hline
			\multicolumn{9}{|l|}{(C = 0) Decuplet $\rightarrow$ Octet}\\\hline
			$\Delta^{+}\rightarrow p$               & $2.480$  & $2.194$  & $2.870$  & $2.47$ & $2.749$  & $2.96$  & $-3.50$ & $2.20(1.10)$  \\
			$\Delta^{0}\rightarrow n$           	& $-2.590$ & $-2.323$ &    -     & $2.47$ & $2.749$  & $2.96$  & $-3.50$ & $-2.20(1.10)$ \\
			$\Sigma^{*0}\rightarrow \Lambda^{0}$	& $2.248$  & $2.017$  & $2.500$  & $2.21$ & $2.380$  & $2.71$  & $3.62$  & $-1.70(1.10)$ \\
			$\Sigma^{*+}\rightarrow \Sigma^{+}$     & $2.138$  & $1.893$  & $2.260$  & $2.21$ & $-2.287$ & $2.76$  & $4.46$  & $1.50(0.60)$  \\
			$\Sigma^{*0}\rightarrow \Sigma^{0}$  	& $0.964$  & $0.824$  & $0.850$  & $0.83$ & $-0.913$ & $1.24$  & $-2.34$ & $-0.65(0.28)$  \\
			$\Sigma^{*-}\rightarrow \Sigma^{-}$ 	& $-0.213$ & $-0.246$ & $-0.550$ & $0.28$ & $0.462$  & $-0.28$ & $-0.21$ & $-0.23(0.07)$ \\
			$\Xi^{*0}\rightarrow \Xi^{0}$       	& $-2.290$ & $-1.985$ & $2.120$  & $2.27$ & $-2.287$ & $2.50$  & $5.38$  & $-1.50(0.50)$ \\
			$\Xi^{*-}\rightarrow \Xi^{-}$       	& $0.315$  & $0.399$  & $-0.470$ & $0.29$ & $0.462$  & $-0.31$ & $0.20$ & $0.20(0.07)$ \\ \hline 
		\end{tabular}
	\end{table}
	%%%%%%%%%%%%%%%%%%%%%%%%%%%%%%%%%%%%%%%%%%%%%%%%%%%%%%%%%%%%%%%%%%%%%%%%%%%%
	\begin{table}
		\centering
		\captionof{table}{Transition moments of charm baryons (in $\mu_N$).} 
		\label{transition_charm}
		\begin{tabular}{|c|c|c|c|c|c|c|c|c|}     \hline
			\multirow{2}{*}{\textbf{Transitions}} & \multirow{2}{*}{\textbf{EMS}} & \multicolumn{2}{c|}{\textbf{SQCS}} &
			\textbf{BM} &\textbf{$\chi$CQM} & \textbf{$\chi$QSM} & \textbf{HB$\chi$PT} & \textbf{LCQSR}\symbolfootnote[12]{Aliev \textit{et al.} have given their results in natural magneton ($e\hbar/2cM_B$); however, to convert to nuclear magneton ($\mu_N$), we multiply the entire magnetic moments with $2m_N/(M_{B_{3/2^+}} + M_{B_{1/2^+}})$.} \\
			\cline{3-4} 
			& & \textbf{$z =0.021$}& \textbf{$z =0.155$}& \bf\cite{Simonis:2018rld} & \bf\cite{Sharma:2010vv} & \bf\cite{Yang:2019tst} & \bf\cite{Wang:2018cre, Li:2017pxa} & \bf\cite{Aliev:2009jt, Aliev:2014bma} \\    \hline    \hline
			\multicolumn{9}{|l|}{(C = 1) Octet $\rightarrow$  Octet}\\\hline
			$\Sigma_c^{+}\rightarrow \Lambda_c^{+}$         & $-1.649$ & $-1.480$ & $-1.480$ & $-1.480$ &  $1.560$ & $1.54(6)$  & $-1.38$ & -  \\
			$\Xi_c^{\prime \,+}\rightarrow \Xi_c^{+}$       & $-1.425$ & $-1.268$ & $-1.317$ & $-1.380$ & $1.300$  & $-1.19(6)$ & $0.73$  & -  \\
			$\Xi_c^{\prime \,0}\rightarrow \Xi_c^{0}$       & $0.182$  & $0.198$  & $0.154$ & $0.139$  & $-0.310$ & $0.21(3)$  & $0.22$  & -  \\    \hline
			\multicolumn{9}{|l|}{(C = 1) Sextet $\rightarrow$  Anti-triplet}\\\hline
			$\Sigma_c^{*+} \rightarrow\Lambda_c^{+}$        & $2.284$  & $2.050$  & $2.050$ & $2.070$  & $2.400$  &$-2.18(8)$ & $2.00$  & $1.48(55)$  \\
			$\Xi_c^{*+} \rightarrow\Xi_c^{+}$               & $1.976$  & $1.758$  & $1.824$ & $1.970$  & $2.080$  &$1.69(8)$  & $1.05$  & $1.47(66)$ \\
			$\Xi_c^{*0} \rightarrow\Xi_c^{0}$               & $-0.249$ & $-0.272$ & $-0.211$ & $-0.193$ & $-0.500$ &$-0.29(4)$ & $-0.31$ & $0.16(7)$ \\\hline
			\multicolumn{9}{|l|}{(C = 1) Sextet $\rightarrow$  Sextet}\\\hline
			$\Sigma_c^{*++} \rightarrow\Sigma_c^{++}$       & $1.180$  & $1.356$  & $1.461$  & $1.340$  & $-1.370$ &$1.52(7)$  & $1.07$  & $1.06(38))$ \\
			$\Sigma_c^{*+} \rightarrow\Sigma_c^{+}$         & $0.029$  & $0.096$  & $0.274$  & $0.102$  & $-0.003$ &$0.33(2)$  & $0.19$  & $0.45(11)$ \\
			$\Sigma_c^{*0} \rightarrow\Sigma_c^{0}$         & $-1.126$ & $-1.165$ & $-0.912$ & $-1.140$ & $1.480$  &$-0.87(3)$ & $-0.69$ & $0.19(8)$ \\
			$\Xi_c^{*+} \rightarrow\Xi_c^{\prime+}$         & $0.159$  & $0.202$  & $0.349$  & $0.216$  & $-0.230$ &$0.43(2)$  &$0.23$   & $0.25(7)$ \\
			$\Xi_c^{*0} \rightarrow\Xi_c^{\prime0}$         & $-1.016$ & $-1.068$ & $-0.845$ & $-1.030$ & $1.240$  &$-0.74(3)$ & $-0.59$ & $0.69(20)$  \\
			$\Omega_c^{*0} \rightarrow\Omega_c^{0}$         & $-0.900$ & $-0.942$ & $-0.751$ & $-0.892$ & $0.960$  &$-0.60(4)$ & $-0.49$ & $0.46(13)$  \\    \hline
			\multicolumn{9}{|l|}{(C = 2) Triplet $\rightarrow$  Triplet}\\\hline
			$\Xi_{cc}^{*++} \rightarrow\Xi_{cc}^{++}$       & $-1.304$ & $-1.359$ & $-1.702$ & $-1.210$ & $1.330$  &  -        & $-2.35$ & -  \\
			$\Xi_{cc}^{*+} \rightarrow\Xi_{cc}^{+}$         & $1.182$  & $1.116$  & $0.700$  & $1.070$  & $-1.410$ &  -        & $1.55$  & -  \\
			$\Omega_{cc}^{*+} \rightarrow\Omega_{cc}^{+}$   & $0.910$  & $0.867$  & $0.596$  & $0.869$  & $-0.890$ &  -        & $1.54$  & -  \\    \hline  
		\end{tabular}
	\end{table}
	%%%%%%%%%%%%%%%%%%%%%%%%%%%%%%%%%%%%%%%%%%%%%%%%%%%%%%%%%%%%%%%%%%%%%
	\begin{table}
		\centering
		\captionof{table}{Radiative M1 decay widths of light baryons (in KeV).} 
		\label{deacy_width_light}
		\begin{tabular}{|c|c|c|c|c|c|c|c|} \hline  
			\textbf{Transitions}\symbolfootnote[13]{The experimental value for $\Gamma_{\Sigma^{*0}\rightarrow \Lambda^{0}\gamma}$, and $\Gamma_{\Sigma^{*+}\rightarrow \Sigma^{+}\gamma}$ obtained by CLAS collaboration are $445(80)$ KeV \cite{CLAS:2011vzg}, and $250(70)$ Kev \cite{CLAS:2011iuw}, respectively.} & \textbf{EMS} & \textbf{SQCS} & \textbf{$\chi$QM \cite{Yu:2006sc}} & \textbf{HCM \cite{Ghalenovi:2017fvw}} & \textbf{CSQM \cite{Ramalho:2020tnn}} & \textbf{$\chi$PT \cite{Li:2017vmq}} & \textbf{LCQSR \cite{Aliev:2004ju}} \\  \hline\hline
			\multicolumn{8}{|l|}{(C = 0) Octet $\rightarrow$ Octet}\\\hline
			$\Sigma^{0}\rightarrow \Lambda^{0}\gamma$    & $9.972$ & $8.021$ & -      & -      & -      & -       & - \\\hline
			\multicolumn{8}{|l|}{(C = 0) Decuplet $\rightarrow$ Octet}\\\hline
			$\Delta^{+}\rightarrow p\gamma$              & $453.3$ & $354.9$ & $363$  & $648$  & $648$  & $730$   & $900(730)$\\
			$\Delta^{0}\rightarrow n\gamma$              & $473.0$ & $380.6$ & $363$  & $648$  & $648$  & $730$   & $900(730)$ \\
			$\Sigma^{*0}\rightarrow \Lambda^{0}\gamma$   & $297.3$ & $239.2$ & $241$  & $325$  & $399$  & $430$   & $470(410)$ \\
			$\Sigma^{*+}\rightarrow \Sigma^{+}\gamma$    & $110.3$ & $86.47$ & $100$  & $149$  & $154$  & $250$   & $110(820)$\\
			$\Sigma^{*0}\rightarrow \Sigma^{0}\gamma$    & $21.66$ & $15.84$ & $16.0$ & $21.0$ & $32.0$ & $70.0$  & $21.0(15.0)$\\
			$\Sigma^{*-}\rightarrow \Sigma^{-}\gamma$    & $1.036$ & $1.385$ & $4.10$ & $2.00$ & $1.40$ & $0.58$ & $2.00(1.00)$\\
			$\Xi^{*0}\rightarrow \Xi^{0}\gamma$          & $178.0$ & $133.8$ & $133$  & $158$  & $182$  & $410$   & $140(90.0)$\\
			$\Xi^{*-}\rightarrow \Xi^{-}\gamma$          & $3.220$ & $5.161$ & $5.40$ & $2.00$ & $2.40$ & $0.52$ & $3.00(2.00)$ \\	\hline 
		\end{tabular}
	\end{table}
	%%%%%%%%%%%%%%%%%%%%%%%%%%%%%%%%%%%%%%%%%%%%%%%%%%%%%%%%%%%%%%%%%%%%%%%%%%%%%%%%%%%%

	\begin{table}
		\caption{Radiative M1 decay widths of charm baryons (in KeV).} 
		\label{decay_width_charm}
		\begin{tabular}{|c|c|c|c|c|c|c|c|c|c|}	\hline 
			\multirow{2}{*}{\textbf{Transitions}} & \multirow{2}{*}{\textbf{EMS}} & \multicolumn{2}{c|}{\textbf{SQCS}}  &
			\textbf{BM} & \textbf{HB$\chi$PT} & \textbf{LCQSR}& \textbf{CQM} & \textbf{$\chi$QSM} & \textbf{hCQM}\\
			\cline{3-4}
			& &\textbf{$z =0.021$}& \textbf{$z =0.155$} &\bf\cite{Simonis:2018rld} & \bf\cite{Wang:2018cre, Li:2017pxa} & \bf\cite{Aliev:2009jt, Aliev:2014bma, Aliev:2016xvq} & \bf\cite{Wang:2017kfr} & \bf\cite{Kim:2021xpp} & \bf\cite{Gandhi:2018lez} \\	\hline	\hline
			\multicolumn{10}{|l|}{(C = 1) Octet $\rightarrow$  Octet}\\\hline
			$\Sigma_c^{+}\rightarrow \Lambda_c^{+}\gamma$       & $93.70$ & $75.46$ & $75.54$ & $74.10$ & $65.60$  & $50.0(17.0)$ 	   & $80.60$ & -       & $66.66$         \\
			$\Xi_c^{\prime+}\rightarrow \Xi_c^{+}\gamma$        & $21.29$ & $16.86$ & $18.18$ & $18.60$ & $5.430$  & $8.50(2.50)$   & $42.30$ & -       & -               \\
			$\Xi_c^{\prime0}\rightarrow \Xi_c^{0}\gamma$        & $0.327$ & $0.389$ & $0.233$ & $0.185$ & $0.460$  & $0.27(6)$    & $0.000$ & -       & -               \\ \hline
			\multicolumn{10}{|l|}{(C = 1) Sextet $\rightarrow$  Anti-triplet}\\\hline
			$\Sigma_c^{*+}\rightarrow \Lambda_c^{+}\gamma$      & $231.6$ & $186.6$ & $187.0$ & $190.0$ & $161.6$  & $130(45)$    & $373.0$ & $69.76$ & $135.3 $       \\
			$\Xi_c^{*+}\rightarrow \Xi_c^{+}\gamma$             & $81.58$ & $64.58$ & $69.50$ & $81.60$ & $21.60$  & $52.0(25.0)$ 	   & $139.0$  & $31.97$ & $15.69 $       \\
			$\Xi_c^{*0}\rightarrow \Xi_c^{0}\gamma$             & $1.263$ & $1.503$ & $0.902$ & $0.745$ & $1.840$  & $0.66(32)$   & $0.000$ & $0.080$ & $0.811 $       \\ \hline
			\multicolumn{10}{|l|}{(C = 1) Sextet $\rightarrow$  Sextet}\\\hline
			$\Sigma_c^{*++}\rightarrow \Sigma_c^{++}\gamma$     & $1.487$ & $1.964$ & $2.278$ & $1.960$ & $1.200$  & $2.65(1.20)$ & $3.940$ & $1.080$ & $2.060$        \\
			$\Sigma_c^{*+}\rightarrow \Sigma_c^{+}\gamma$       & $0.001$ & $0.010$ & $0.081$ & $0.011$ & $0.040$  & $0.40(16)$   & $0.004$ & $0.060$ & $4 \t 10^{-5}$ \\
			$\Sigma_c^{*0}\rightarrow \Sigma_c^{0}\gamma$       & $1.370$ & $1.468$ & $0.900$ & $1.410$ & $0.490$  & $0.08(3)$    & $3.430$ & $0.300$ & $2.162 $       \\
			$\Xi_c^{*+}\rightarrow \Xi_c^{\prime+}\gamma$       & $0.030$ & $0.049$ & $0.146$ & $0.063$ & $0.070$  & $0.27$     & $0.004$ & $0.090$ & -              \\
			$\Xi_c^{*0}\rightarrow \Xi_c^{\prime0}\gamma$       & $1.263$ & $1.396$ & $0.875$ & $1.330$ & $0.420$  & $2.14$      & $3.030$ & $0.340$ & -              \\
			$\Omega_c^{*0}\rightarrow \Omega_c^{0}\gamma$       & $1.142$ & $1.250$ & $0.790$ & $1.130$ & $0.320$  & $0.93$      & $0.890$ & $0.340$ & $0.464$        \\ \hline
			\multicolumn{10}{|l|}{(C = 2) Triplet $\rightarrow$  Triplet}\\\hline
			$\Xi_{cc}^{*++}\rightarrow \Xi_{cc}^{++}\gamma$     & $2.390$ & $2.595$ & $4.070$ & $2.790$ & $22.00$  & -            & -       & -       & -              \\
			$\Xi_{cc}^{*+}\rightarrow \Xi_{cc}^{+}\gamma$       & $1.963$ & $1.752$ & $0.689$ & $2.170$ & $9.570$  & -            & -       & -       & -   			\\
			$\Omega_{cc}^{*+}\rightarrow \Omega_{cc}^{+}\gamma$ & $1.969$ & $1.789$ & $0.844$ & $1.600$ & $9.450$  & -            & -       & -       & -   			 \\ \hline
		\end{tabular}
	\end{table}

	%%%%%%%%%%%%%%%%%%%%%%%%%%%%%%%%%%%%%%%%%%%%%%%%%%%%%%%%%%%%%%%%%%%%%%%%%%5
	\begin{table}
		\centering
		\captionof{table}{State mixing in baryon magnetic (transition) moments (in $\mu_N$).} 
		\label{mixing_mm}
		\begin{tabular}{|c|c|c|c|c|c|c|c|c|} \hline
			\textbf{Quark} & \textbf{Mixing angle} & \multirow{2}{*}{\textbf{Baryon}} &\textbf{EMS} & \textbf{SQCS} & \textbf{EMS} & \textbf{SQCS} & \textbf{BM}& \textbf{NRQM}\\
			\textbf{content} & \textbf{(degree)} & & \textbf{mixed} & \textbf{mixed} & \textbf{unmixed} & \textbf{unmixed} & \textbf{mixed \cite{Simonis:2018rld}} & \textbf{mixed \cite{Simonis:2018rld, Bernotas:2012nz}}\\ \hline	\hline
			&          & $\Lambda^{0}$                                    & $-0.604$ & $-0.524$ & $-0.579$ & $-0.502$ & -        & -     \\
			&          & $\Sigma^{0}$                                     & $0.850$  & $0.720$  & $0.825$  & $0.698$  & -        & -     \\
			$uds$ & $0.412$ & $\Sigma^{0}\rightarrow \Lambda^{0}$         & $-1.696$ & $-1.522$ & $-1.706$ & $-1.530$ & -        & -     \\
			&          & $\Sigma^{*0}\rightarrow \Lambda^{0}$             & $2.255$  & $2.023$  & $2.248$  & $2.017$  & -        & -     \\
			&          & $\Sigma^{*0}\rightarrow \Sigma^{0}$              & $0.947$  & $0.809$  & $0.964$  & $0.824$  & -        & -     \\   \hline
			
			&          & $\Lambda_{c}^{+}$                                & $0.377$  & $0.381$  & $0.380$  & $0.384$  & -        & $0.390$\\
			&          & $\Sigma_{c}^{+}$                                 & $0.435$  & $0.537$  & $0.432$  & $0.534$  & -        & $0.490$ \\
			$udc$ & $0.057$ & $\Sigma_{c}^{+}\rightarrow\Lambda_{c}^{+}$  & $-1.649$ & $-1.480$ &$-1.649$  & $-1.480$ & -        & $-1.610$\\
			&          & $\Sigma_{c}^{*+}\rightarrow\Lambda_{c}^{+}$      & $2.284$  & $2.050$  & $2.284$  & $2.050$  & -        & $2.200$  \\
			&          & $\Sigma_{c}^{*+}\rightarrow\Sigma_{c}^{+}$       & $0.026$  & $0.094$  & $0.029$  & $0.096$  & -        & $0.070$ \\	\hline
			
			&          & $\Xi_{c}^{+}$                                    & $0.197$  & $0.222$  & $0.380$  & $0.384$  & $0.142$  & $0.200$ \\
			&          & $\Xi_{c}^{\prime\,+}$                            & $0.806$  & $0.850$  & $0.623$  & $0.688$  & $0.825$  & $ 0.890$\\
			$usc$ & $3.707$ & $\Xi_{c}^{\prime\,+}\rightarrow\Xi_{c}^{+}$ & $-1.398$ & $-1.238$ & $-1.425$ & $-1.268$ & $-1.330$ &$-1.400$ \\
			&          & $\Xi_{c}^{*+}\rightarrow\Xi_{c}^{+}$             & $1.982$  & $1.768$  & $1.976$  & $1.758$  & $1.860$  &$ 2.030$ \\
			&          & $\Xi_{c}^{*+}\rightarrow\Xi_{c}^{\prime\,+}$     & $0.031$  & $0.088$  & $0.159$  & $0.202$  & $0.066$  & $0.090$ \\	\hline
			
			&          & $\Xi_{c}^{0}$                                	  & $0.397$  & $0.391$  &  $0.380$ & $0.372$  & $0.346$  & $ 0.410$ \\
			&          & $\Xi_{c}^{\prime\,0}$                            & $-1.091$ & $-1.176$ & $-1.074$ & $-1.157$ & $-1.130$ & $-1.180 $\\
			$dsc$ & $3.650$ & $\Xi_{c}^{\prime\,0}\rightarrow\Xi_{c}^{0}$ & $0.088$  & $0.100$  & $0.182$  & $0.198$  & $0.034$  & $0.080$  \\
			&          & $\Xi_{c}^{*0}\rightarrow\Xi_{c}^{0}$             & $-0.313$ & $-0.339$ & $-0.249$ & $-0.272$ & $-0.249$ & $-0.330$ \\
			&          & $\Xi_{c}^{*0}\rightarrow\Xi_{c}^{\prime\,0}$     & $-0.998$ & $-1.048$ & $-1.016$ & $-1.068$ & $-0.994$ & $-1.070$ \\	\hline 
		\end{tabular}
	\end{table}	
	%%%%%%%%%%%%%%%%%%%%%%%%%%%%%%%%%%%%%%%%%%%%%%%%%%%%%%%%%%%%%%%%%%%%%%%%%%%%%%%%%%%%
	\begin{table}
		\centering
		\captionof{table}{State mixing in radiative M1 decay widths (in KeV).} 
		\label{mixing_decay_width}
		\begin{tabular}{|c|c|c|c|c|c|}	\hline 
			\multirow{2}{*}{\textbf{Transitions}} & \textbf{EMS} & \textbf{SQCS} &\textbf{EMS} &\textbf{SQCS} & \textbf{BM}\\
			& \textbf{mixed} & \textbf{mixed} & \textbf{unmixed} & \textbf{unmixed} & \textbf{mixed \cite{Simonis:2018rld}}\\	\hline	\hline
			$\Sigma^{0}\rightarrow \Lambda^{0}\gamma$          & $9.852$ & $7.929$ & $9.972$ & $8.021$ & -     \\
			$\Sigma^{*0}\rightarrow \Lambda^{0}\gamma$         & $299.1$ & $240.5$ & $297.3$ & $239.2$ & -     \\
			$\Sigma^{*0}\rightarrow \Sigma^{0}\gamma$          & $20.94$ & $15.28$ & $21.66$ & $15.84$ & -     \\ \hline
			$\Sigma_{c}^{+}\rightarrow\Lambda_{c}^{+}\gamma$   & $93.70$ & $75.45$ & $93.70$ & $75.46$ & -     \\
			$\Sigma_{c}^{*+}\rightarrow\Lambda_{c}^{+}\gamma$  & $231.7$ & $186.6$ & $231.6$ & $186.6$ & -      \\
			$\Sigma_{c}^{*+}\rightarrow\Sigma_{c}^{+}\gamma$   & $0.001$ & $0.010$ & $0.001$ & $0.010$ & -      \\ \hline
			$\Xi_{c}^{\prime\,+}\rightarrow\Xi_{c}^{+}\gamma$  & $20.47$ & $16.07$ & $21.29$ & $16.86$ & $17.30$ \\
			$\Xi_{c}^{*+}\rightarrow\Xi_{c}^{+}\gamma$         & $82.09$ & $65.27$ & $81.58$ & $64.58$ & $72.70$ \\
			$\Xi_{c}^{*+}\rightarrow\Xi_{c}^{\prime\,+}\gamma$ & $0.001$ & $0.009$ & $0.030$ & $0.049$ & $0.006$ \\ \hline
			$\Xi_{c}^{\prime\,0}\rightarrow\Xi_{c}^{0}\gamma$  & $0.076$ & $0.098$ & $0.327$ & $0.389$ & $0.011$ \\
			$\Xi_{c}^{*0}\rightarrow\Xi_{c}^{0}\gamma$         & $1.997$ & $2.341$ & $1.263$ & $1.503$ & $1.240$  \\
			$\Xi_{c}^{*0}\rightarrow\Xi_{c}^{\prime\,0}\gamma$ & $1.219$ & $1.346$ & $1.263$ & $1.396$ & $1.230$  \\	
			\hline 
		\end{tabular}
	\end{table}	
	%%%%%%%%%%%%%%%%%%%%%%%%%%%%%%%%%%%%%%%%%%%%%%%%%%%%%%%%%%%%%%%%%%%%%%%%%%%%%%%%%%%%
\end{document}